\documentclass{aa}  

\usepackage{graphicx}
\usepackage{txfonts}
\usepackage{lipsum}
\usepackage{comment}
\usepackage{subcaption}         
\usepackage{lscape}             
\usepackage{placeins}           
\usepackage{enumitem}

\usepackage{txfonts}
\usepackage[colorlinks=true, citecolor=blue, urlcolor=blue]{hyperref}
\usepackage{placeins}
\usepackage{color, soul}
\usepackage{xcolor}
\usepackage{ulem}
\usepackage{natbib}
\usepackage{lipsum}
\bibpunct{(}{)}{;}{a}{}{,} 
\usepackage{tabularx}  

\begin{document}

   \title{Population synthesis of Galactic middle-aged pulsar wind nebulae}

   \subtitle{I.~Detection prospects for current and future instruments}


   \author{A. De Sarkar\inst{1}\thanks{desarkar@ice.csic.es}, D. F. Torres{\inst{1,2,3}}\thanks{dtorres@ice.csic.es},  B. Olmi{\inst{4}},  N. Bucciantini{\inst{4,5,6}}, D. M.-A. Meyer{\inst{1}}
   }

   \institute{Institute of Space Sciences (ICE, CSIC), Campus UAB, Carrer de Can Magrans s/n, 08193 Barcelona, Spain
   \and
   Institut d'Estudis Espacials de Catalunya (IEEC), Gran Capit\`a 2-4, 08034 Barcelona, Spain 
   \and
   Instituci\'o Catalana de Recerca i Estudis Avan\c cats (ICREA), 08010 Barcelona, Spain 
             \and INAF - Osservatorio Astrofisico di Arcetri, Largo E. Fermi 5, I-50125 Firenze, Italy
            \and Università degli Studi di Firenze, Via Sansone 1, 50019, Sesto F.no (Firenze), Italy 
            \and INFN - Sezione di Firenze, Via G. Sansone 1, I-50019 Sesto F.no (Firenze), Italy
}

   \date{Received XXXX XX, 20XX}
 
    \abstract
   {
    Pulsar wind nebulae (PWNe) constitute the largest population of Galactic very high-energy (VHE; $E > 100$ GeV) $\gamma$-ray sources and are key laboratories for studying particle acceleration and pulsar–supernova remnant (SNR) interactions. However, realistic population-level predictions have so far lacked any detailed treatment of the reverberation phase, when the nebula is compressed by the SNR reverse shock, significantly altering its dynamics and radiative spectrum. We employed the hybrid \texttt{TIDE+L} framework, which combines a thin-shell dynamical model with a Lagrangian treatment of the SNR structure during reverberation, allowing for self-consistent evolution of thousands of PWNe across all stages up to $10^5$ yr. Each source was evolved under distributions of pulsar spin-down, SNR, and environmental properties, and the resulting $\gamma$-ray fluxes were used to estimate the detectability by current and next-generation $\gamma$-ray observatories, while accounting for their sensitivity and sky coverage. The model predicts that the upcoming Cherenkov Telescope Array Observatory (CTAO) will detect an order of magnitude more PWNe than those firmly detected in the tera-electronvolt range, confirming its dominant contribution to the forthcoming tera-electronvolt population census. Our results demonstrate that realistic modeling of reverberation is important for predicting the Galactic tera-electronvolt PWNe population. 
    }

   \titlerunning{Population synthesis of middle-aged PWNe - I}
   \authorrunning{De Sarkar et al.}
   \maketitle
   \nolinenumbers

\section{Introduction}

Pulsars, the rapidly rotating and highly magnetized remnants of massive stellar explosions, lose their rotational energy predominantly through the emission of a relativistic wind of particles and magnetic field.
When this wind interacts with the surrounding medium, it forms a pulsar wind nebula (PWN) -- a bubble of relativistic plasma bounded by a termination shock where particles, primarily leptons,
are accelerated to extreme energies.
The PWNe radiate broadband non-thermal emission predominantly via synchrotron and inverse Compton (IC) processes, extending from radio to peta-electronvolt $\gamma$ rays (see \cite{gaensler06, slane17, olmi23} for comprehensive reviews).
Because they probe both the pulsar central engine and the interaction with the host supernova remnant (SNR) and interstellar medium (ISM), PWNe are unique laboratories for high-energy astrophysics, relativistic plasma physics, and particle acceleration under extreme conditions.  

The PWNe constitute the most numerous class of firmly identified very high-energy (VHE, $E > 100$ GeV) Galactic $\gamma$-ray sources detected by current instruments such as High Energy Stereoscopic System (H.E.S.S.) \citep{hess18b, hess18}, Major Atmospheric Gamma-ray Imaging Cherenkov (MAGIC) \citep{magic10, rico16}, and Very Energetic Radiation Imaging Telescope Array System (VERITAS) \citep{veritas08, aliu13, aliu14, mukherjee16}.
They are also key targets for wide-field detectors such as High-Altitude Water Cherenkov (HAWC) and the Large High Altitude Air Shower Observatory (LHAASO) \citep{hawc20, lhaaso24}.
Understanding their global properties, demographics, evolution, and radiative output is therefore essential both for interpreting current surveys and for guiding the expectations of, for example, the upcoming Cherenkov Telescope Array Observatory (CTAO; see, e.g., \citep{wilhelmi13}) and Southern Wide-field Gamma-ray Observatory (SWGO; 
see, e.g., \citet{conceicao23}).

The evolution of PWNe is intimately tied to the dynamical stage of its host SNR, 
itself a function of the past stellar wind history of its progenitor and of the local conditions 
of the ISM. 
In the early free-expansion phase, the nebula expands within the cold, homologously expanding ejecta, and its size and energetics closely track the pulsar’s spin-down luminosity. 
However, the majority of Galactic PWNe are not young systems: they are middle-aged, mostly beyond $\sim 10^{4}$ yr. Early on in this stage, the SNR reverse shock travels back toward the explosion center and collides with the nebula. 
This event initiates the reverberation phase, in which the PWN is compressed, heated, and later {re-expands} within the surrounding ejecta \citep{reynolds84, gelfand09, torres18, bandiera23III}. 
Its effects are profound: the nebular magnetic field can be amplified by orders of magnitude, particle populations can be reenergized, and radiative losses are drastically enhanced. 
These processes strongly reshape the PWN broadband spectrum and temporal evolution.

Despite its importance, reverberation has remained the most challenging phase of PWN evolution to model. 
On the one hand, one-zone thin-shell models have been widely employed in population studies and spectral fittings, owing to their computational simplicity and speed \citep[e.g.][]{pacini73, reynolds84, becker07, zhang08, qiao09, gelfand09, fang10, tanaka10, tanaka11, bucciantini11, martin12, Vorster13, torres13, torres14, martin16, martin22,vanrensburg18, zhu18, torres18, torres19,fiori20, desarkar22,kolb17,temim15}. 
In the best of cases, these models approximate the swept-up shell as infinitely thin and impose analytic prescriptions for the confining SNR. 
While robust during the free-expansion phase $-$ when the SNR ejecta {are} cold and self-similar $-$ the reliability of these assumptions rapidly decreases during reverberation.
At this stage, the interaction between the PWN and the SNR’s reverse shock becomes highly nonlinear, involving complex processes such as shell thickening, shock reflections, and dynamical feedback that lie beyond the reach of simple analytical prescriptions.
On the other hand, multidimensional hydrodynamic (HD) and magnetohydrodynamic (MHD) simulations can accurately follow the complex interplay of shocks, turbulence, and magnetic fields \citep{blondin01,vanderswaluw01, bucciantini03, komissarov04, delzanna06, porth14, olmi16, olmi20} and have recently included detailed descriptions of the circumstellar medium \citep[CSM;][]{meyer22,meyer24,meyer25a,meyer25}. 
However, relativistic {MHD simulations are prohibitively expensive for a long PWN evolution}: they typically consume millions of CPU hours to model only a few hundred years and have constraints in volume too.
As a result, they cannot be readily applied to the long-term evolution of hundreds or thousands of PWNe, as required for statistical studies or for comparison with survey data.

This modeling gap and large computational cost  has motivated the development of the hybrid \texttt{TIDE+L} framework \citep{bandiera20, bandiera21, bandiera23II, bandiera23III}. 
\texttt{TIDE+L} combines the efficiency of the thin-shell approach in the free-expansion phase with a Lagrangian treatment of the SNR structure during reverberation. 
This strategy allows the PWN--SNR system to be evolved self-consistently across all stages, including the complex compression episodes of reverberation, while also following the evolution of the particle spectrum subject to radiative and adiabatic losses. 
The code is fast enough to simulate thousands of systems, yet accurate enough to capture the essential physics of reverberation -- at least when compared with 1D HD simulations.
This means that \texttt{TIDE+L} represents a step forward for population synthesis studies of PWNe connecting pulsar birth properties to the observable $\gamma$-ray sky.  
The need for such an approach is particularly acute when addressing detection prospects. 
Because middle-aged PWNe dominate the Galactic population, and because their spectra and luminosities are heavily shaped by reverberation, predictions that neglect this phase risk {achieving biased} conclusions.
Accurate forecasts are necessary to interpret present source catalogs, assess selection effects, and anticipate the yield of surveys with CTAO, SWGO, LHAASO, and other observatories.

In this work, we present our efforts to achieve population synthesis of Galactic middle-aged PWNe, including a more realistic treatment of reverberation. 
We constructed a synthetic population of PWNe sampled from distributions of the initial spin period, magnetic field, ejecta mass, and other SNR and ambient properties, and evolved each system with \texttt{TIDE+L} up to a maximum age of $10^{5}$ yr. 
After computing the radiative and dynamical evolution of each PWN, we compared the results with the flux sensitivity thresholds, while taking into account the sky coverage of current and upcoming facilities (H.E.S.S., HAWC, CTAO, LHAASO, and {SWGO}) {as well as the sensitivity degradation effect due to source extension}, thereby quantifying, in a statistically robust way, the fraction of the Galactic PWNe population that would be detectable.
This paper is the first of a two-part series. Here, in Paper I, we examine the detectability of Galactic PWNe and establish the statistical baseline for their visibility in $\gamma$-ray surveys.
Paper II will focus on the most extreme outcome of reverberation -- superefficiency -- where the nebular luminosity can temporarily exceed the pulsar’s spin-down power.
Together, we aim for these papers to provide a unified framework for understanding the Galactic PWN population, its evolutionary diversity, and its observational signatures in the era of next-generation $\gamma$-ray astronomy.

This paper is organized as follows. 
In Sect. \ref{pop}, we describe the underlying distributions that define the initial population and briefly outline the method used to evolve it. 
Sect. \ref{results} presents and discusses the main results of the population synthesis, while Sect. \ref{conclusion} summarizes our conclusions.

\section{The population and its evolution}\label{pop}

\subsection{Number of sources in the population}

The synthetic population of Galactic PWNe was generated to reproduce the statistical properties of middle-aged systems, while remaining consistent with pulsar birth distributions and core-collapse supernova (SN) progenitors. 
In total, $\sim 2400$ sources were simulated.
Out of the total number of sources, we randomly considered $1600$ sources, 
corresponding to the expected number of PWNe formed in the Galaxy, assuming a core-collapse SN rate of {about} $1.6$ {per century}, reported as the best-fit value in \cite{rozwa21}, implying the formation of about $1600$ PWNe over the past $10^{5} \ \mathrm{yrs}$.
We neglected systems older than $10^{5}$~yr, which are expected to be too diluted to contribute significantly. 
This SN explosion rate is a factor of $\sim$3-4 larger than recent progenitor-based estimates of the core-collapse SN rate derived from near-complete censuses of OB stars in the solar neighborhood \citep{Quintana2025}. 
However, such lower values would be difficult to reconcile with the observed population of young PWNe, unless as-yet-unknown systematic biases in pulsar or PWN population modeling, such as tera-electronvolt detectability times or radiative efficiencies, significantly affect current estimations.

To account for the uncertainty resulting from individual population realizations,
{we performed 1000 random realizations of samples containing 1600 sources each}, drawn from our full population. The ensemble of realizations provides the mean expected number of detectable PWNe, as well as the associated uncertainties arising purely from population stochasticity and sampling variance.

\begin{figure*}[t!]
\centering
\includegraphics[width=\columnwidth]{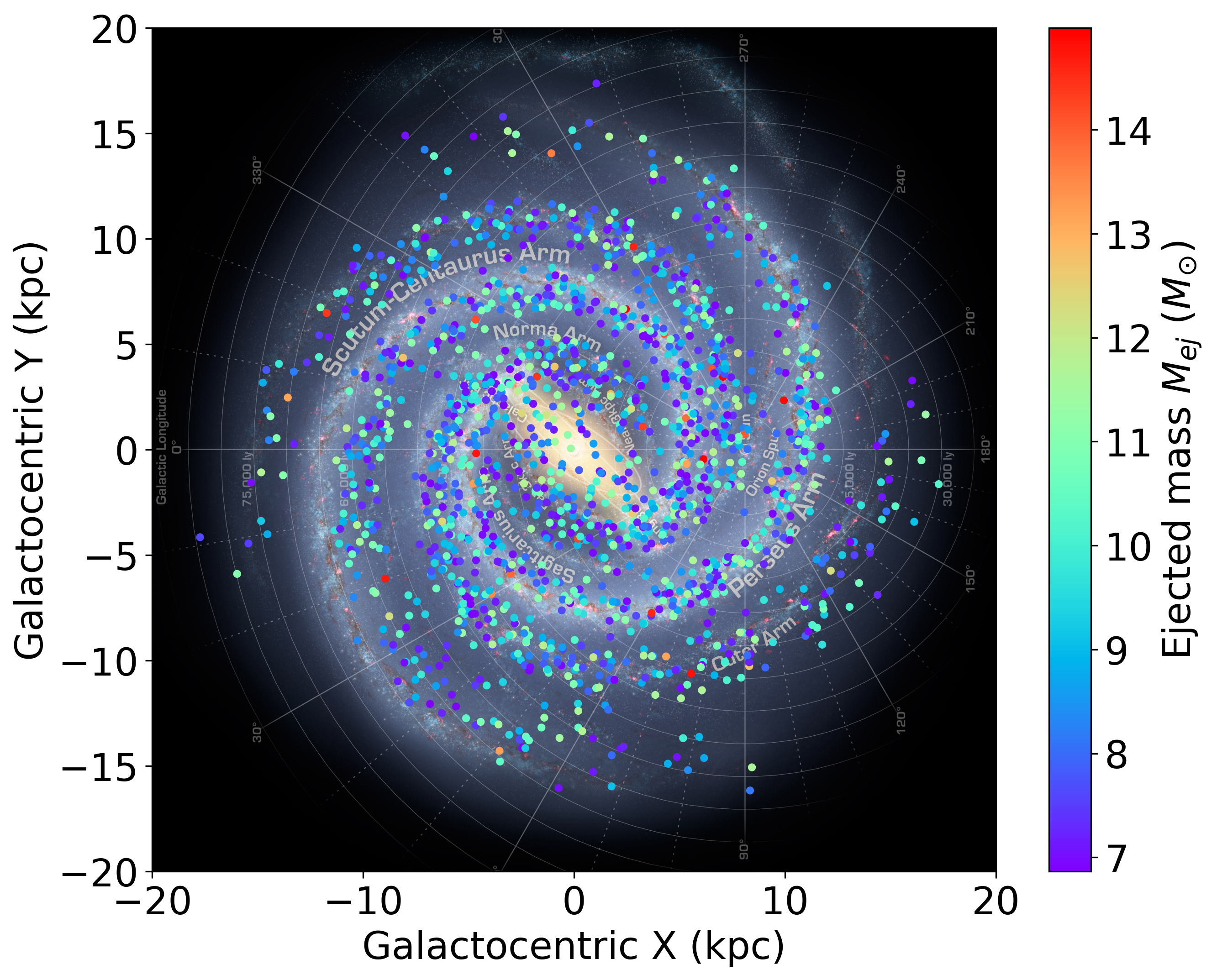}
\raisebox{1.7cm}{\includegraphics[width=\columnwidth]{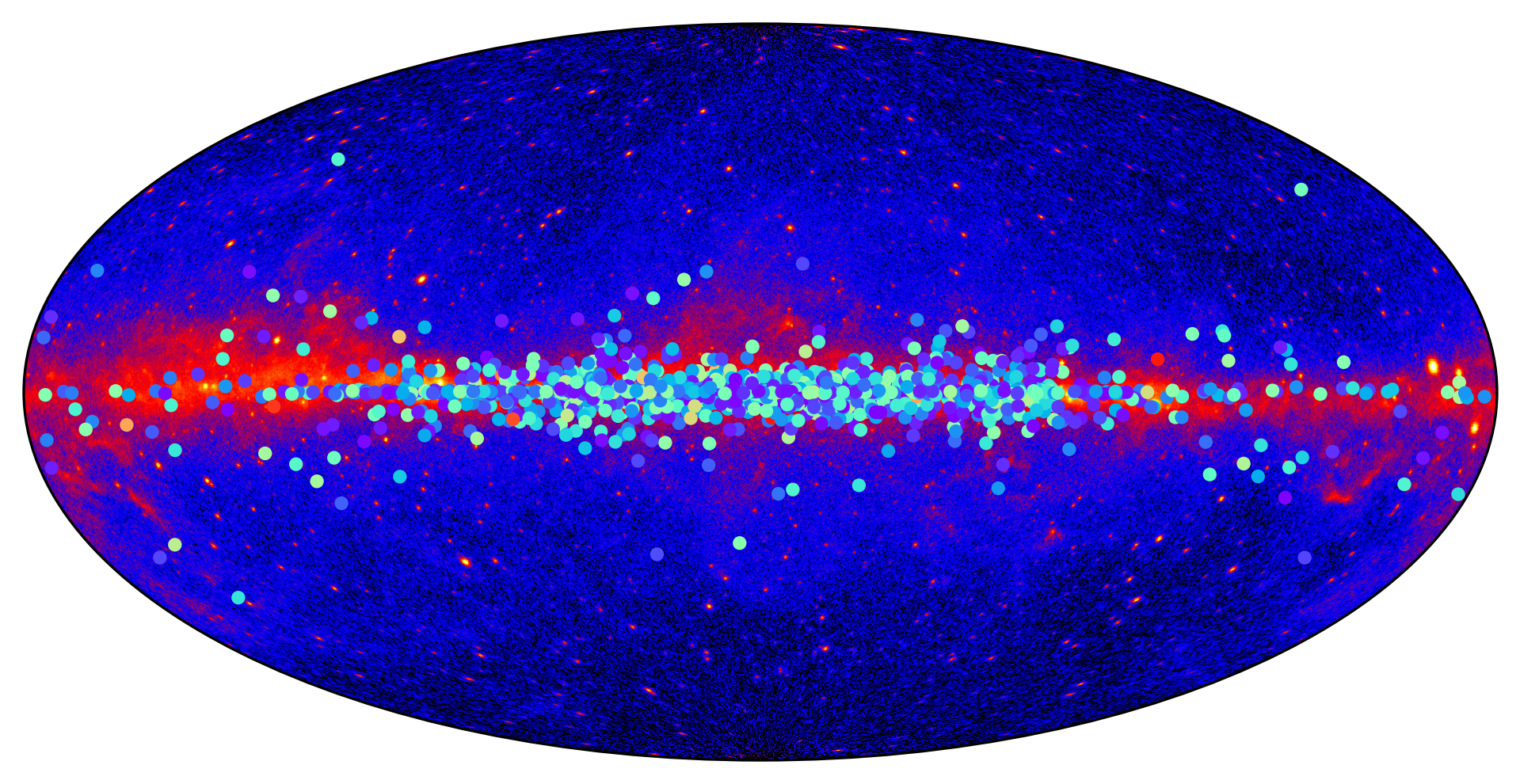}}
\caption{Spatial distribution of the synthetic PWN population in the Milky Way Galaxy for {a random} realization of $1600$ sources. The corresponding ejecta masses (in units of M$_{\odot}$) of each PWN are color-coded as indicated in the color bar. The {right panel} of the figure shows the {\it Fermi}-LAT $\gamma$-ray skymap and the distribution of the same PWNe population with respect to it. The background top-down (face-on) view of the Milky Way (Credit: NASA/JPL-Caltech/R. Hurt (SSC/Caltech)) and the all-sky $\gamma$-ray (Aitoff) skymap (Credit: NASA/DOE/Fermi LAT Collaboration) were generated using the \texttt{mw-plot} code \citep{mwplot}. 
}
\label{fig: mwplot}
\end{figure*}
 
\begin{figure*}
\centering
\includegraphics[width=0.33\linewidth]{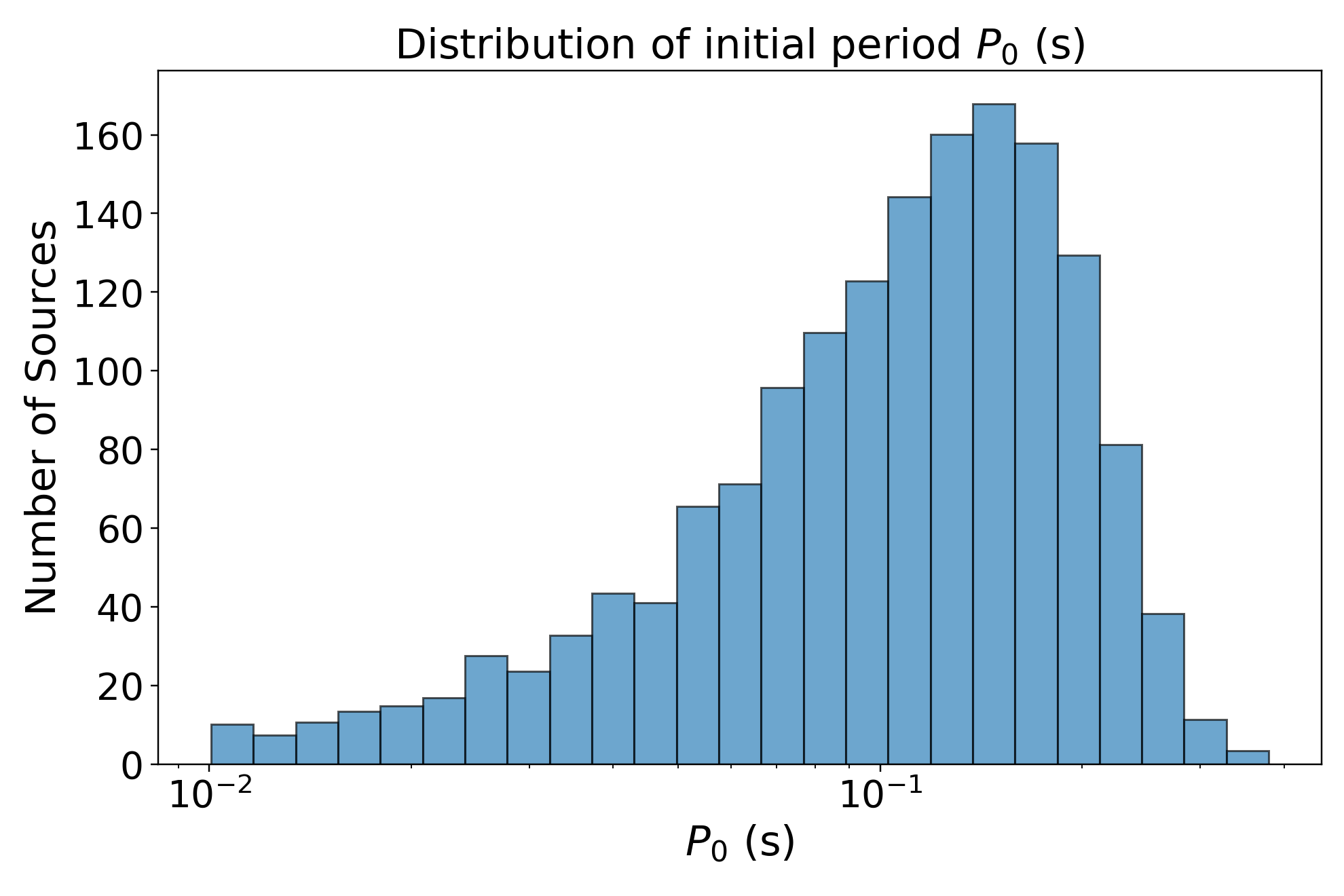}
\includegraphics[width=0.33\linewidth]{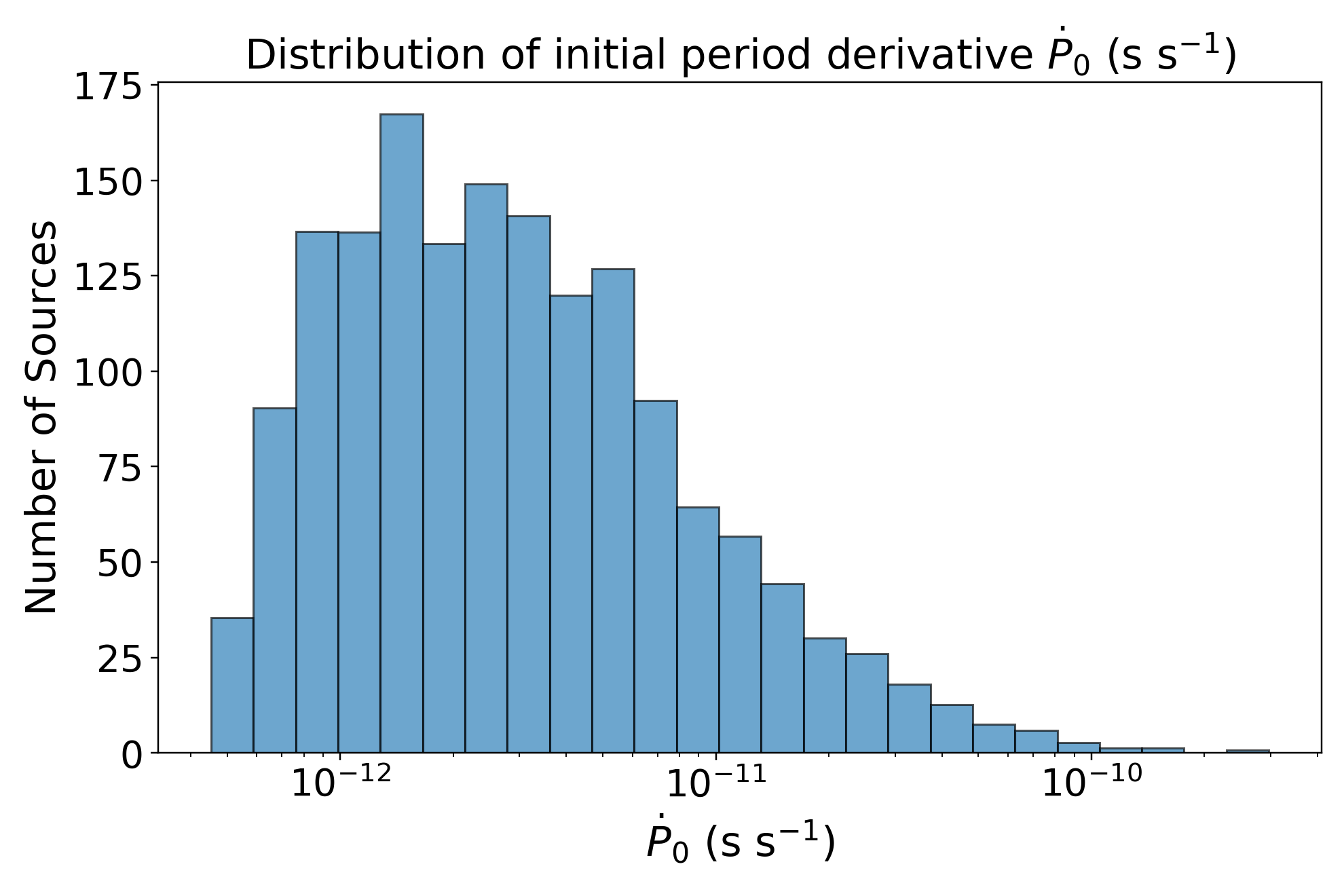}
\includegraphics[width=0.33\textwidth]{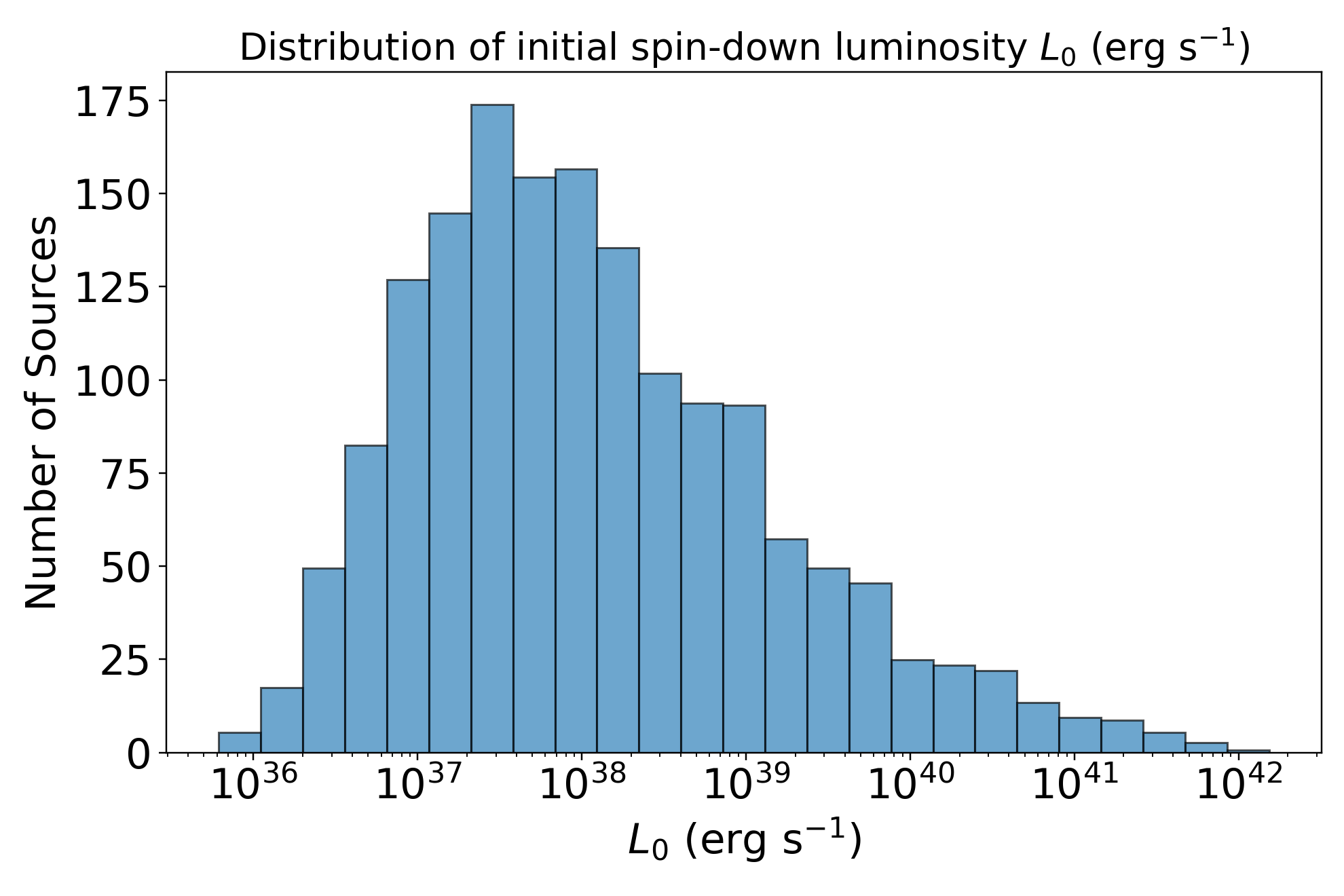}
\includegraphics[width=0.33\textwidth]{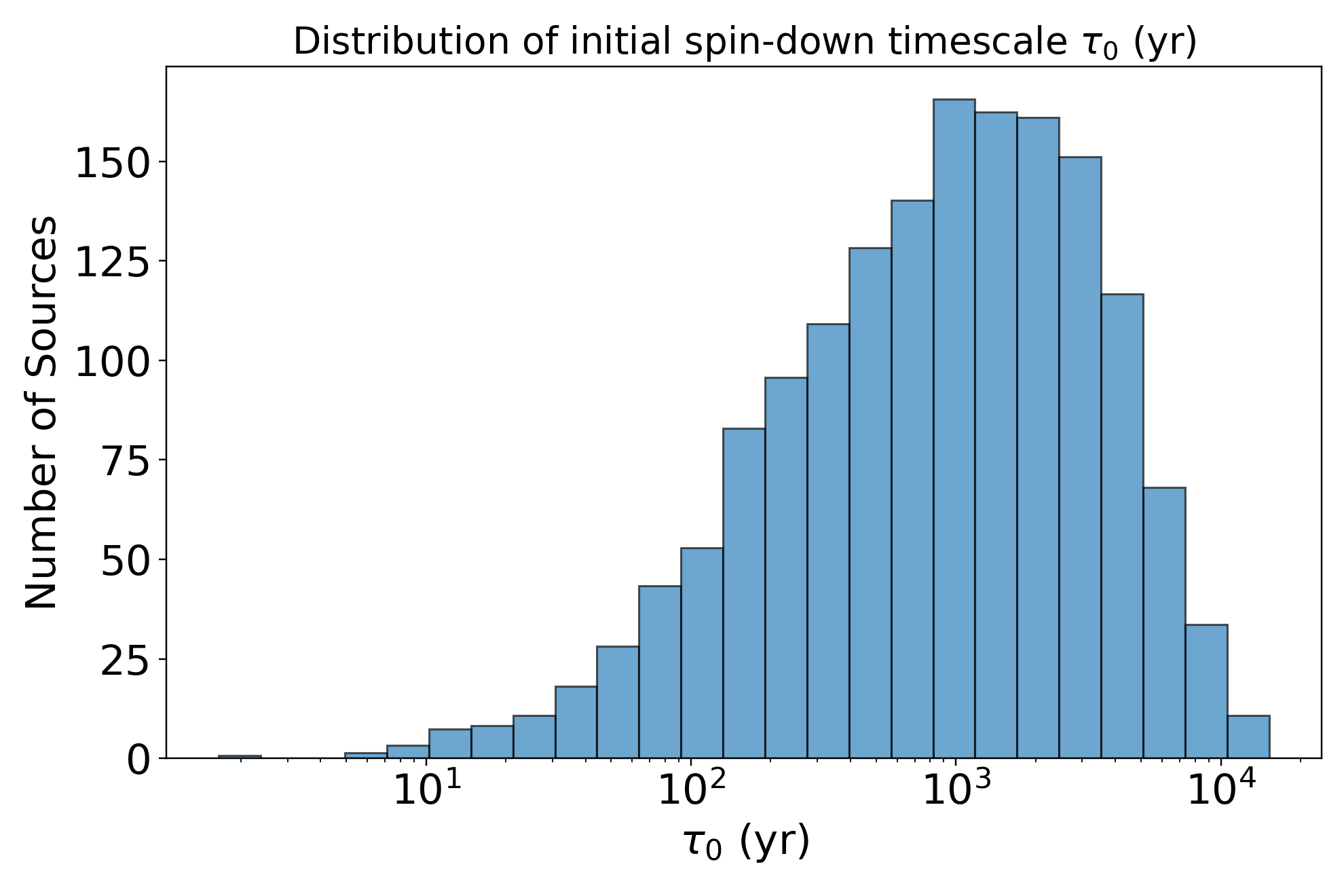}
\includegraphics[width=0.33\linewidth]{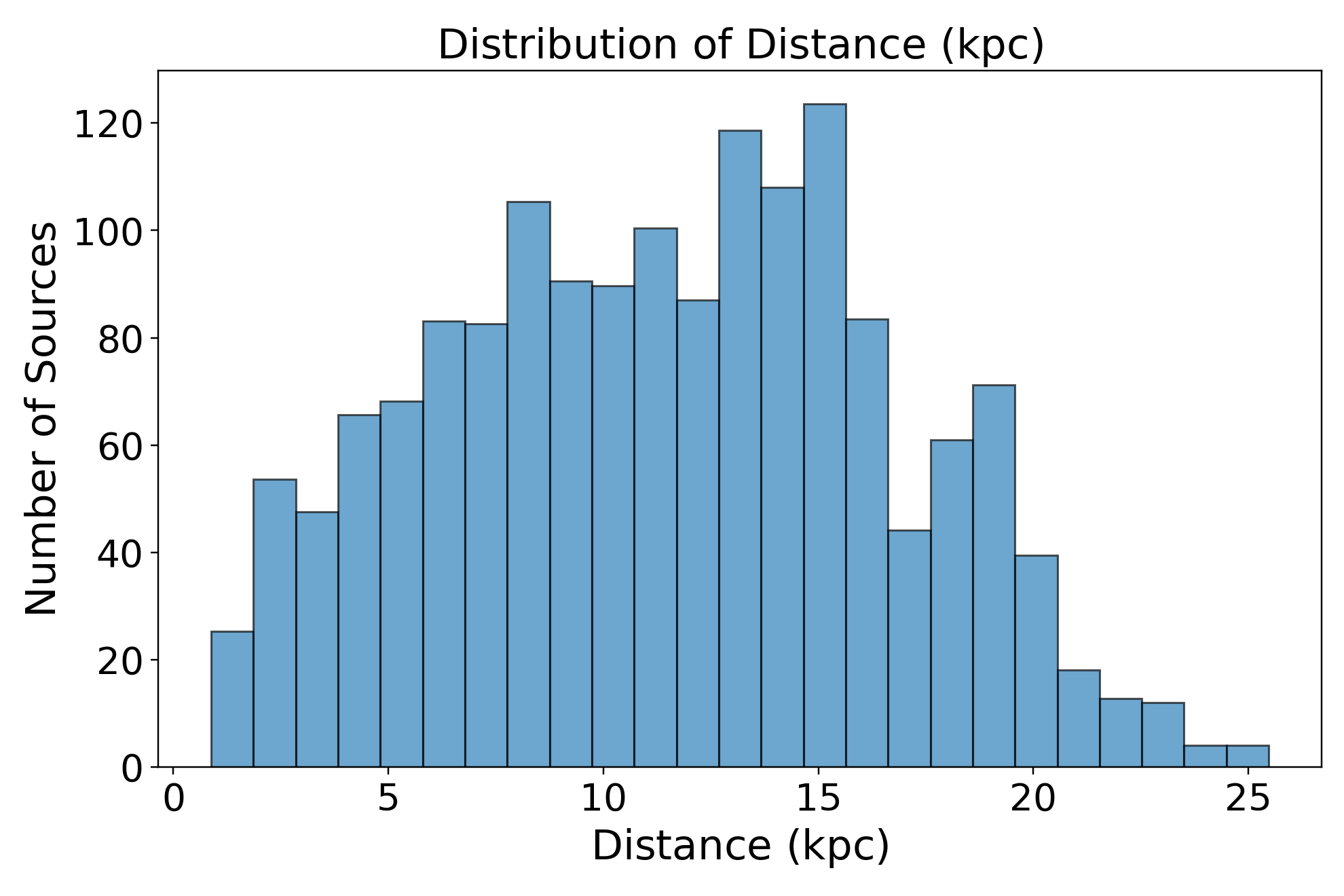}
\includegraphics[width=0.33\linewidth]{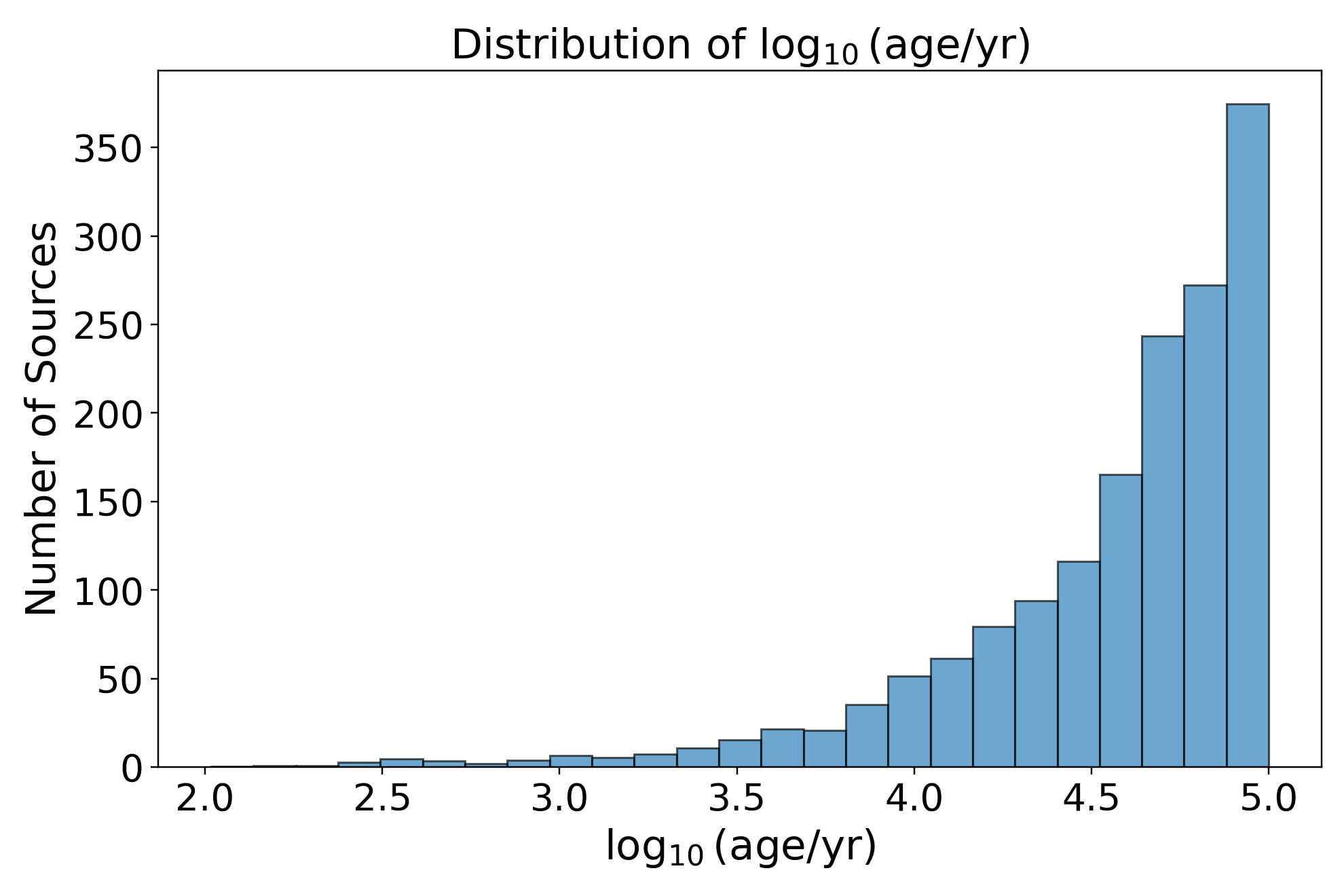}
\includegraphics[width=0.33\linewidth]{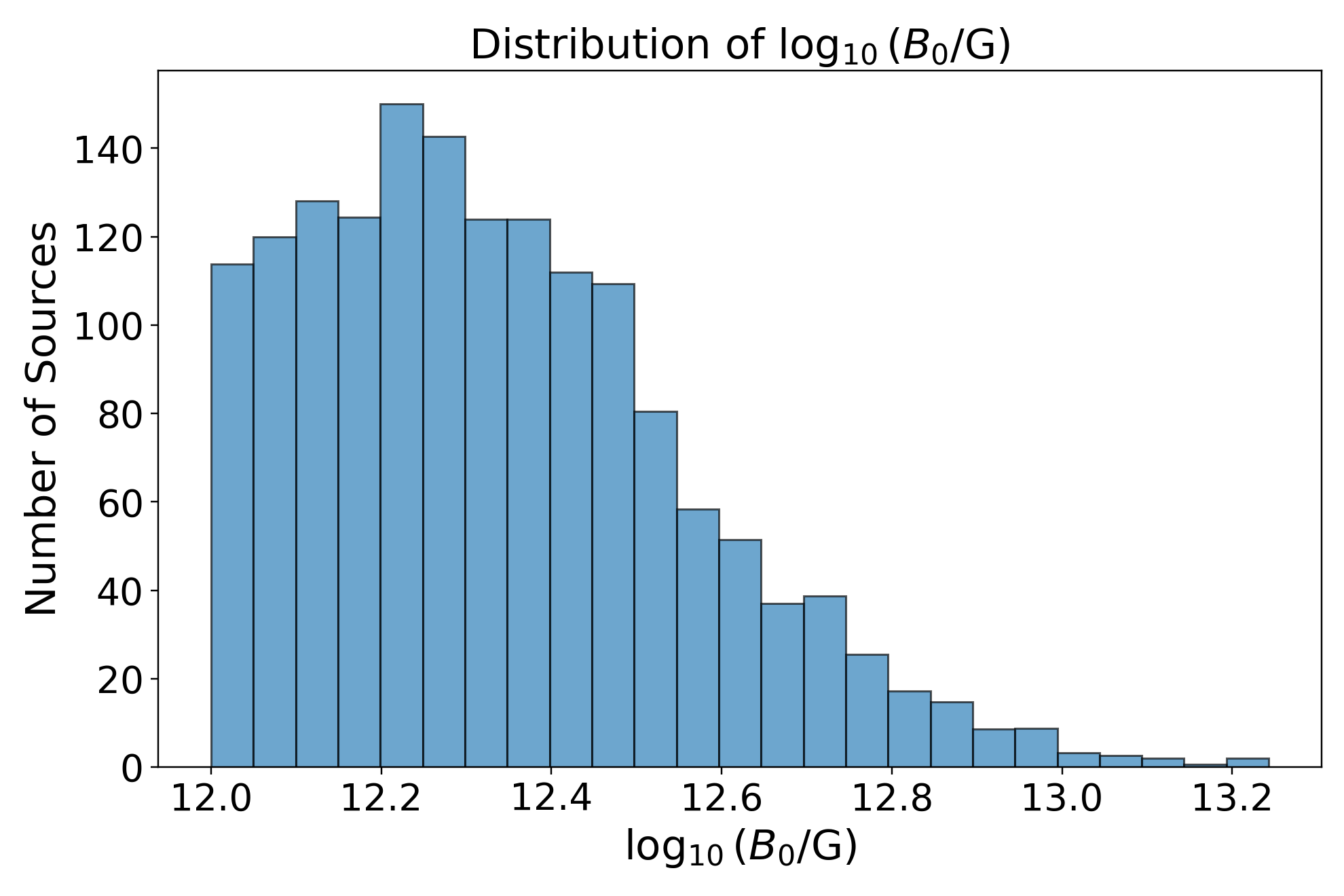}
\includegraphics[width=0.33\linewidth]{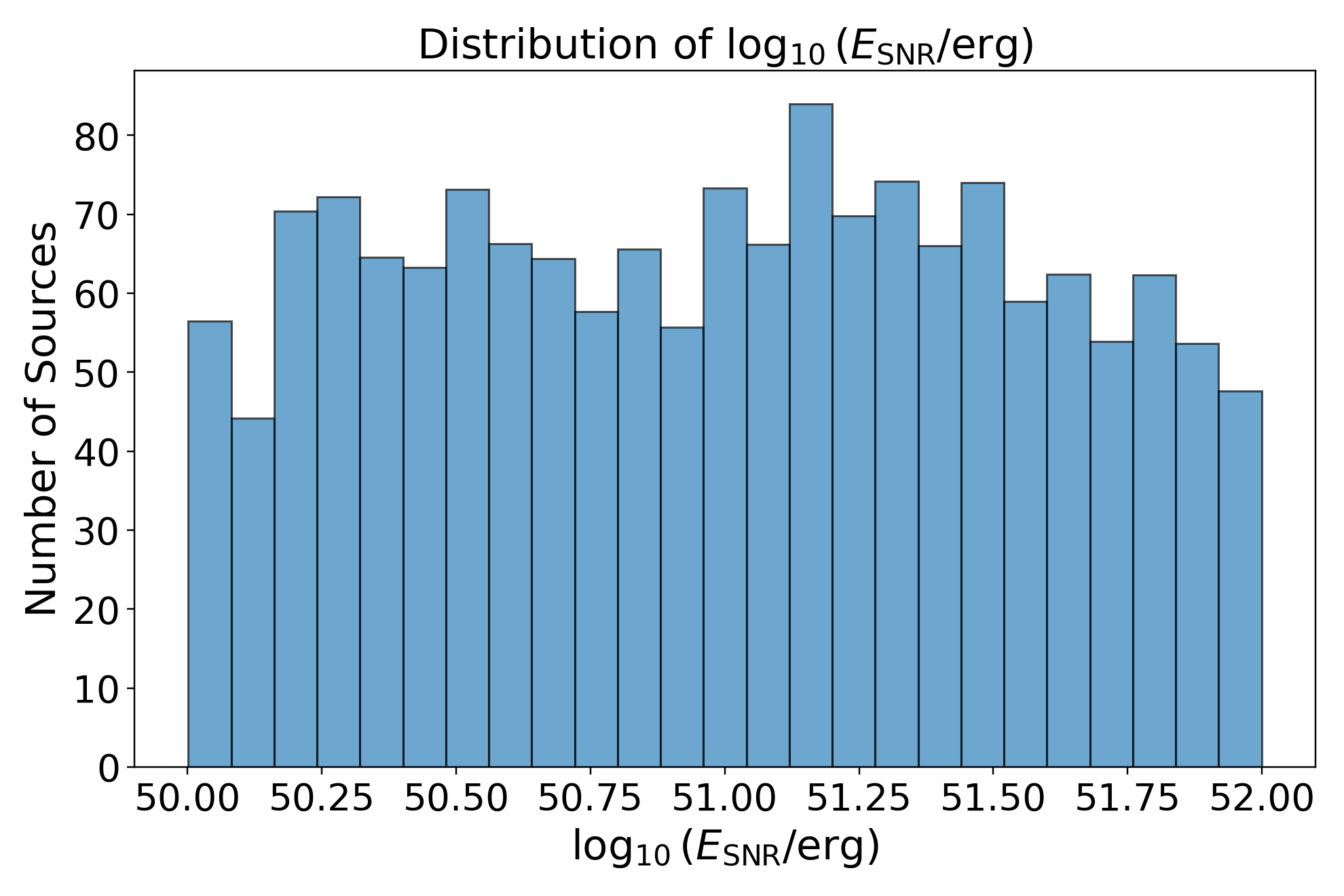}
\includegraphics[width=0.33\linewidth]{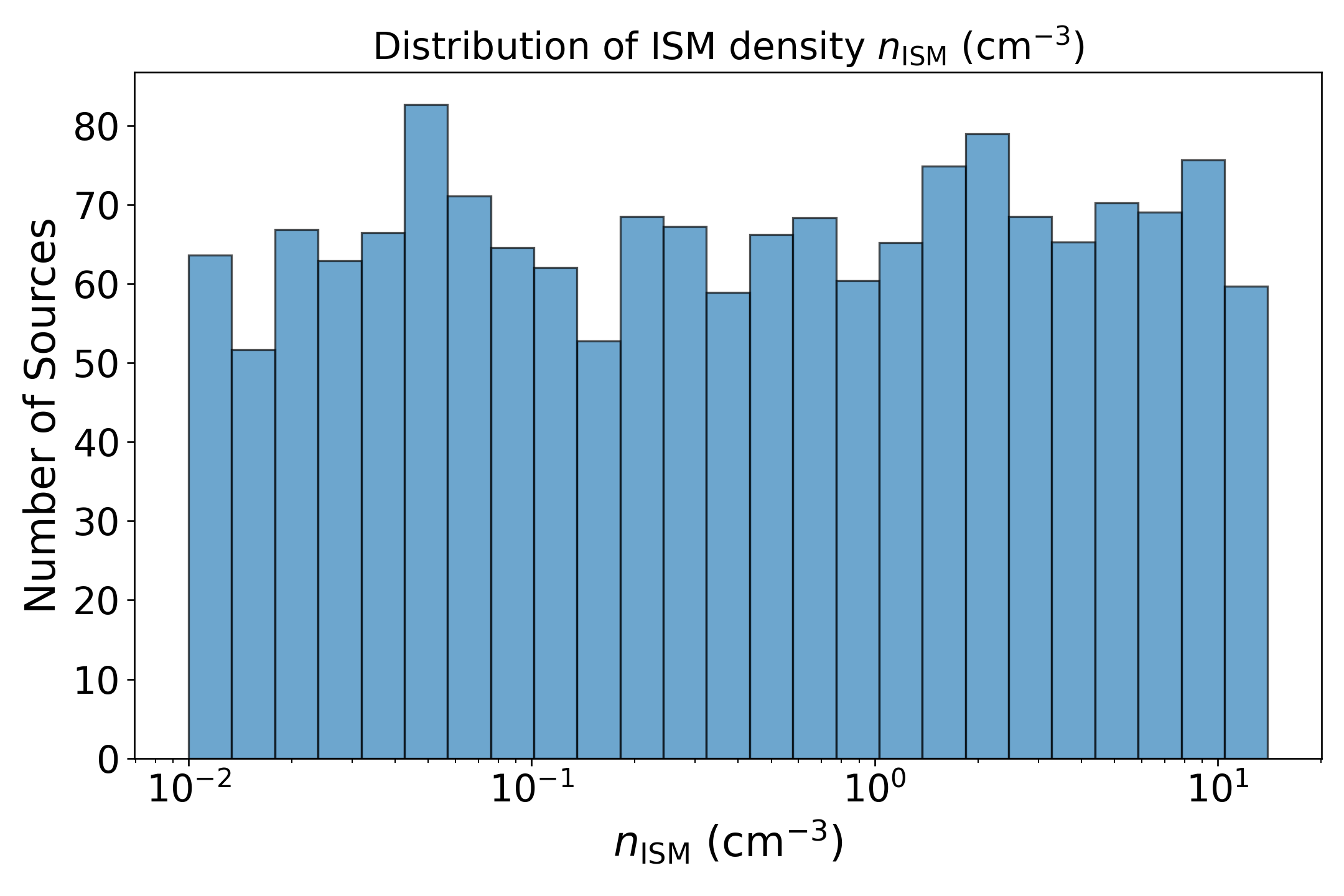}
\includegraphics[width=0.33\textwidth]{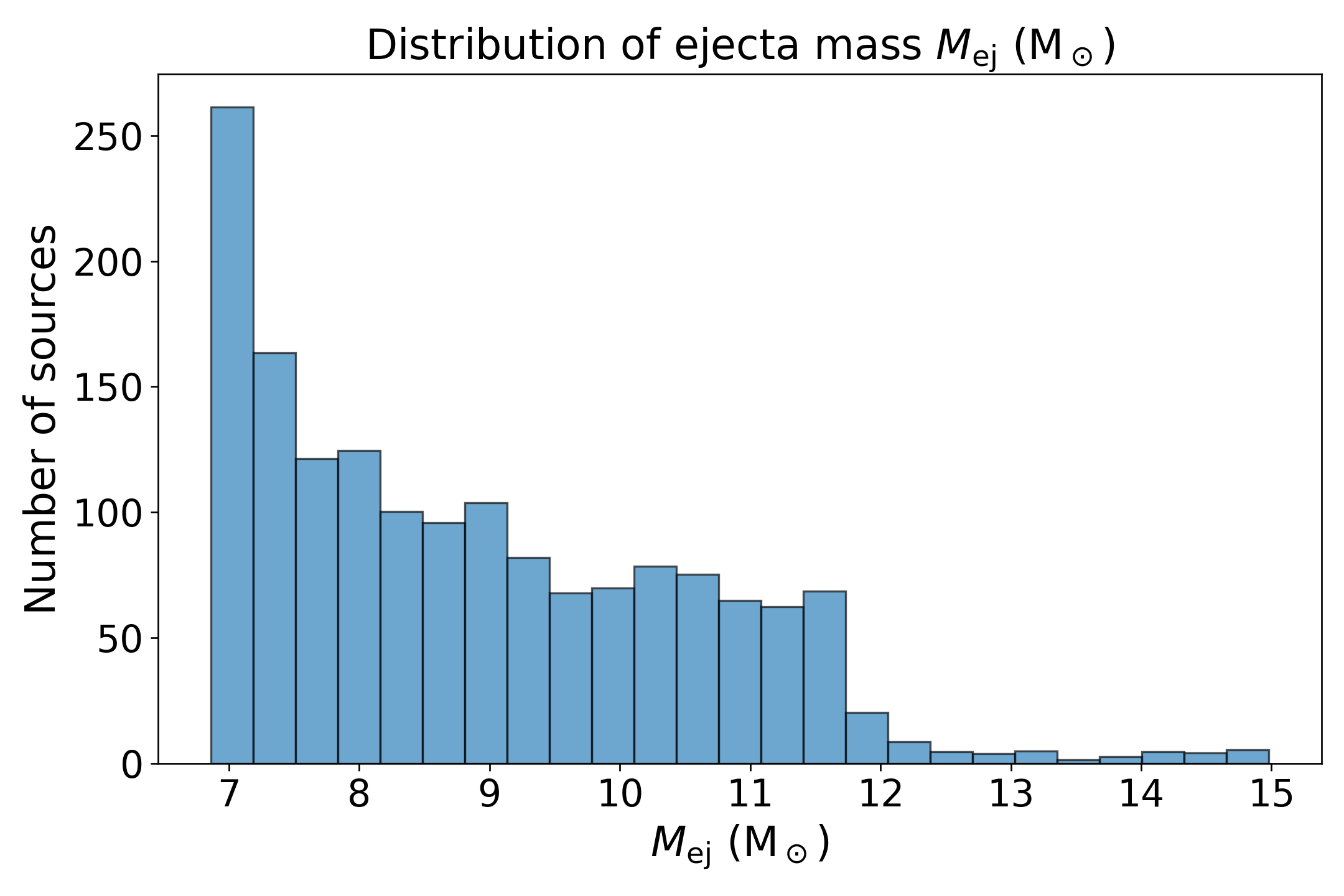}
\includegraphics[width=0.33\textwidth]{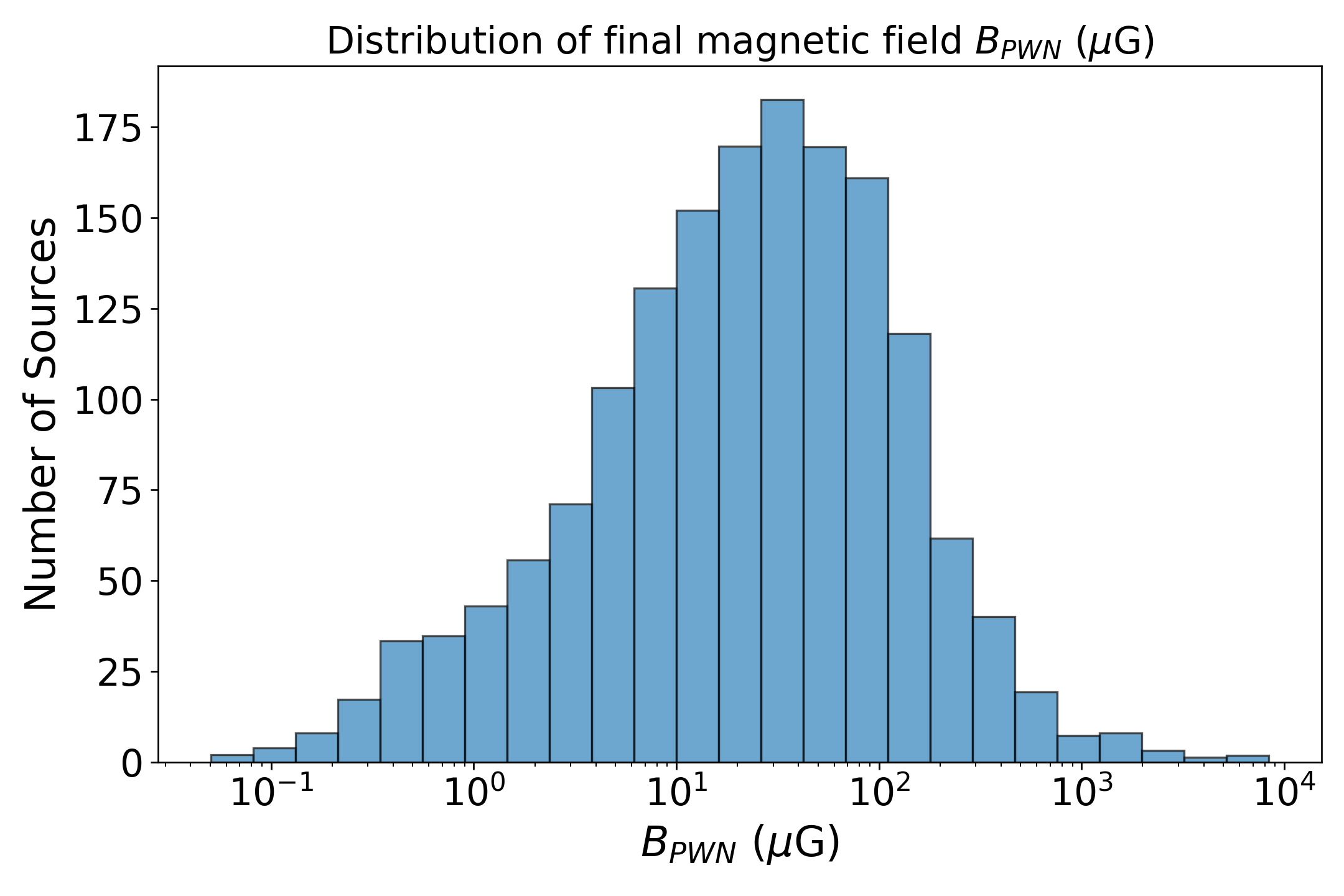}
\caption{Histograms of the parameter distributions employed in this work. {Each histogram shows the mean number of sources per bin, obtained by averaging the bin counts over 1000 random realizations, each built by drawing 1600 PWNe without replacement from the full synthetic parent population of 2400 systems.} The bin-by-bin uncertainty corresponds to the 1$\sigma$ scatter among realizations, and typically ranges from $\approx 7\%$ to $\approx 15\%$ for the parameters shown. {Note that the final magnetic field distribution, $B_{\rm PWN}$, shown in the figure was not assumed a priori, but instead emerges from the simulation. The histogram was constructed using the PWN magnetic field evaluated at the current age of each source.}}
\label{fig: parameter_distribution}
\end{figure*}

\begin{figure}[t!]
\centering
\includegraphics[width=\columnwidth]{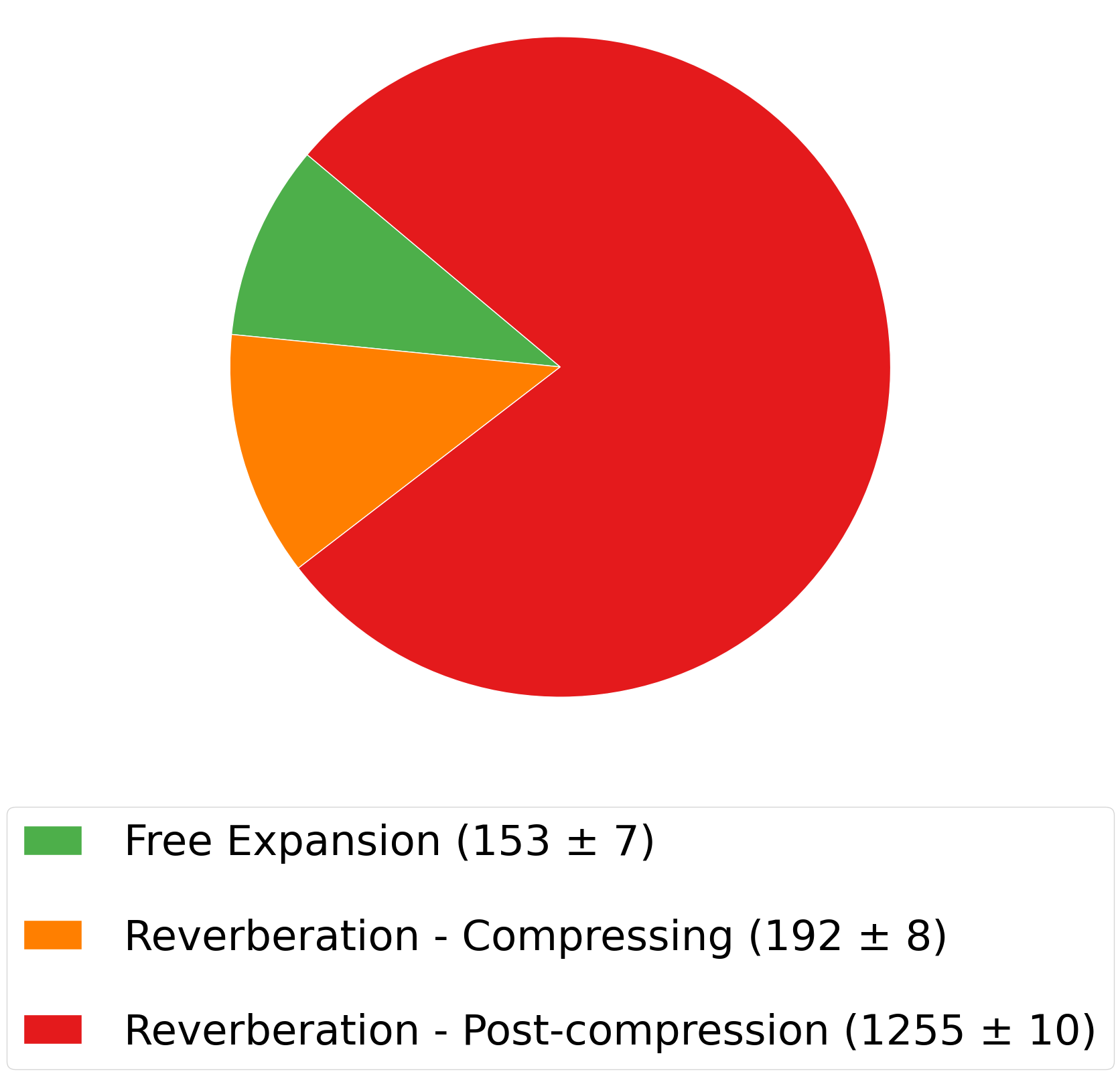}
\caption{Pie chart distinguishing sources in free expansion (green), reverberation in the compressing phase (orange), and reverberation in the post-compression phase (red). The plotted counts correspond to the mean number of sources in each evolutionary stage, obtained by averaging over 1000 realizations of $1600$ randomly drawn PWNe at the final simulation time. The 1$\sigma$ uncertainties for each phase are also provided in the legend.
} 
\label{fig: piechart}
\end{figure}

\begin{figure*}[t!]
\centering
\includegraphics[width=1.01\columnwidth]{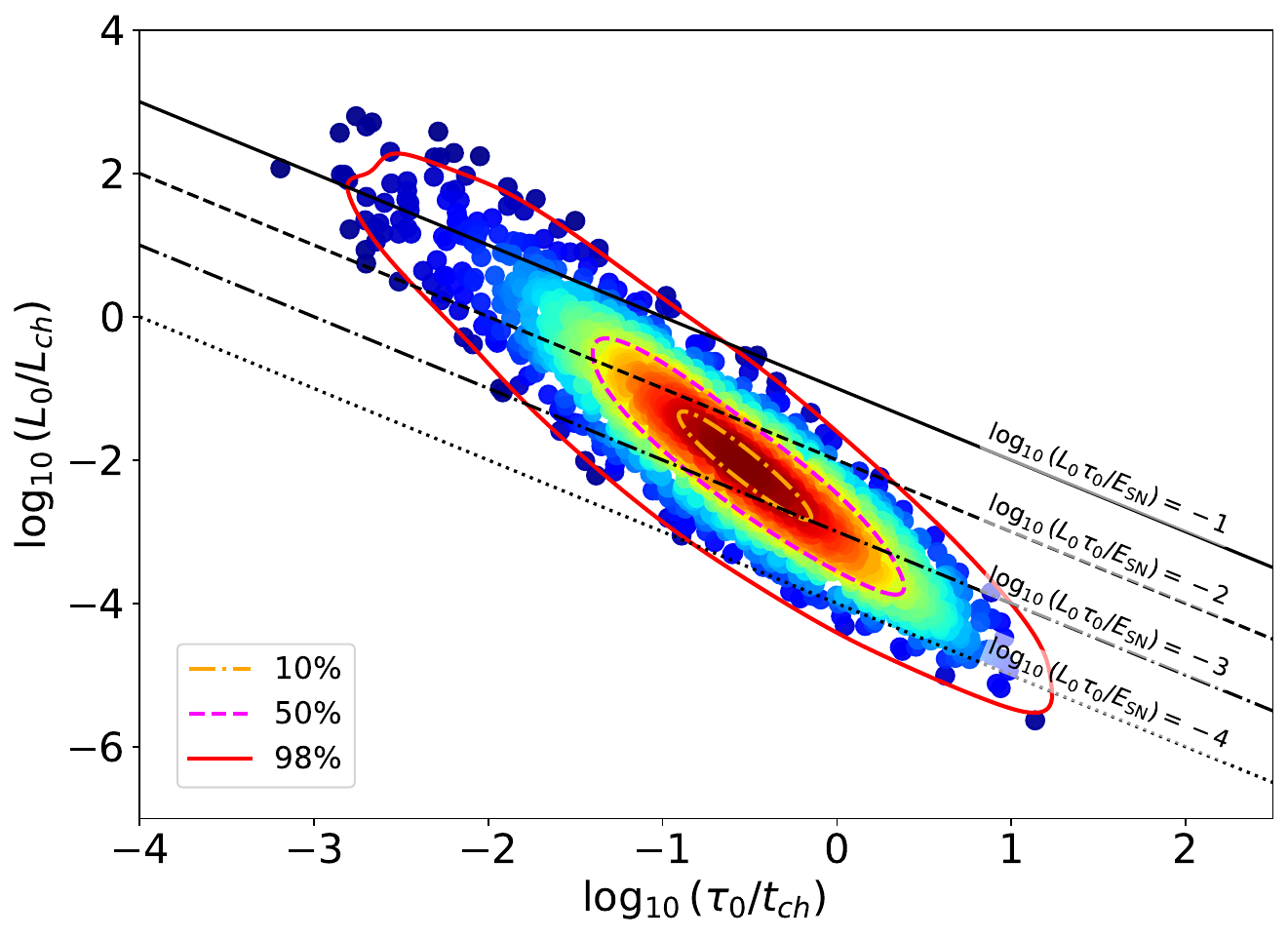}
\raisebox{2.5mm}{\includegraphics[width=\columnwidth]{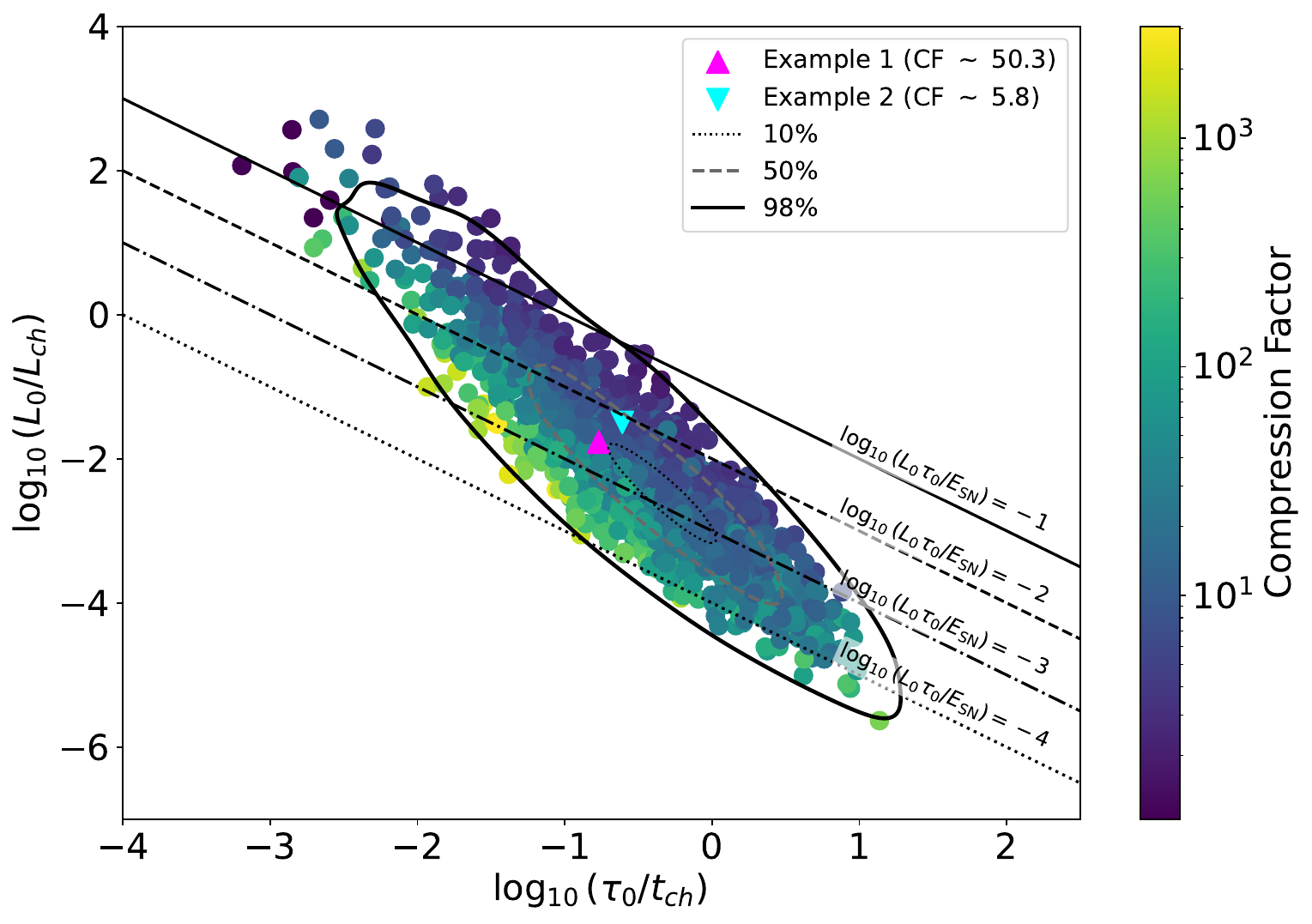}}
\caption{The distribution of the population on the characteristic log$_{10}(L_0/L_{\rm ch})$–log$_{10}(\tau_0/t_{\rm ch})$ plane. The left panel shows a single {random} realization of the synthetic PWN population containing $1600$ sources on the plane, with ellipses marking the isolevels enclosing 10, 50, and 98 percent of the population. Each point is color-coded by density, derived from Gaussian KDE, with colors ranging from red (denser regions) to blue (sparser ones). The right panel displays the same plane along with the isolevels, but color-coded by the CF (CF = $R_{\max}/R_{\min}$) at the first compression, as indicated in the color bar. In this plot, only PWNe in the post-compression phase are shown. Colored triangles in this plot highlight two example PWNe {within a 50$\%$ enclosing ellipse} with different CFs used for individual simulations. Straight lines in both plots denote loci of four constant pulsar–SNR energetics.}
\label{fig: isocontour}
\end{figure*}

\subsection{Spatial distribution}

For the spatial distribution of the synthetic PWNe population, we used the same approach as is presented in \cite{cristofari17}, whereby the core collapse SNRs are positioned in the Galaxy according to the spatial distribution of the Galactic pulsar population as modeled by \cite{fgk06}.
A similar approach was adopted by \cite{fiori22}, as the method was optimized to reproduce the observed population of Galactic $\gamma$-ray SNRs.
Distances to each of the PWNe were assigned to mimic the large-scale pulsar distribution \citep{fiori22}, ensuring a spread across longitude and latitude consistent with observed populations.

\subsection{Age distribution}

The age distribution of the population was randomly sampled from the SN birth rate, with a cut excluding systems younger than 100 years. This was assumed to maintain observational consistency, as no confirmed nearby SN explosion has been observed in the last 100 years, and any such nearby event would likely be detected unless severely obscured by dust.

\subsection{SN explosion energy distribution}

The explosion energy is typically considered to be fixed at $E_{\rm SN} = 10^{51}$ erg; however, in reality, the core collapse SN energetics show variability \citep{hamuy02, nadyozhin03, zampieri03, muller17}.
To account for this variation, the distribution of SN explosion energy was calculated using a truncated Gaussian distribution. 
After imposing lower and upper truncation limits at $10^{50}$ and $10^{52}\,\mathrm{erg}$, respectively, the considered truncated distribution has a mean $\langle \log_{10}(E_{\rm SNR}/{\rm erg}) \rangle = 51$ and a standard deviation of $0.54$ \citep{batzofin24,batzofin25, martinez22}.

\subsection{Interstellar medium characteristics}

The ISM density $n_{\rm ISM} (=\rho_{\rm ISM} / 1.27 \ m_p)$ follows a uniform distribution in the range of $0.01 \leq n_{\rm ISM} \leq 10\ {\rm \,cm^{-3}}$.
{Note that the factor of 1.27 accounts for a standard chemical composition including hydrogen, helium, and traces of heavier elements.}

Our range of ISM densities is different to the values assumed by \cite{cristofari17}, in which the authors consider a wider range ($10^{-5} \leq n_{\rm ISM} \leq 10\ {\rm \,cm^{-3}}$).
A reasonable minimum cutoff was implemented in the ISM distribution according to expectations of typical Galactic environments, and to maintain the characteristic radius of the underlying SNR within a physically consistent range. 
As an example, for a choice of $n_{\rm ISM} = 10^{-5} {\rm \,cm^{-3}}$, the SNR characteristic radius would be around 200-400 pc, depending on the value of the ejecta mass, which is inconsistent with the properties of currently observed Galactic SNRs. 
Such low-density environments may, in principle, host physically large, extremely radio-faint, diluted remnants that fall below the sensitivity of current surveys, thereby easing the tension regarding the total number of SNRs expected based on standard ccSN rates.
{Nevertheless, in this paper, we adopt a lower cutoff in $n_{\rm ISM}$ at $0.01 \ \rm cm^{-3}$ to probe tenuous environments, while avoiding the production of overly diluted SNRs with unrealistically large characteristic radii, for which there is currently no clear observational support in Galactic SNR catalogs.}

\subsection{Estimation of ejecta mass from progenitors}

To determine the ejecta mass, $M_{\rm ej}$, we relied on the relation between the zero-age main-sequence (ZAMS) mass and ejecta mass for nonrotating core-collapse SN progenitors at solar metallicity, provided by tabulated models 
generated by the GENEVA stellar evolution code \citep{eggenberger2008,ekstrom12}. 
GENEVA\footnote{https://www.unige.ch/sciences/astro/evolution/en/database} simulates stellar structures 
in a one dimensional fashion, taking into account the microphysical processes 
of convection and diffusion in the stellar interior, using the Schwarzschild mixing length 
criterion as well as the Zahn viscosity diffusion coefficient for angular momentum transport. 
The physics of the mass-loss rate is described in \citet{georgy2021}.
This relation, illustrated in Fig. 1 of \cite{batzofin24}, was used to map each progenitor ZAMS mass to its corresponding $M_{\rm ej}$.
To do this, the ZAMS masses were sampled from an initial mass function (IMF), which represents the distribution of stellar masses at birth. 
For high-mass stars, the IMF is well approximated by a power law of the form $p(M) \propto M^{-\alpha}$, where $\alpha = 2.35$ corresponds to the Salpeter slope. 
We considered stellar masses in the range $M \in [8, 50]\, \rm M_\odot$.
The higher end is unlikely to end in neutron stars (instead of black holes) 
but it is used here so that there is the possibility of finding a resulting maximum ejecta mass (from GENEVA tracks) at higher values. With a maximum ZAMS mass of $25\, \rm M_\odot$, the ejecta results in around 12 $\rm M_{\odot}$.
Since a higher ejecta mass has been estimated from already-established PWNe, we have increased it to 50 $\rm M_{\odot}$ to include slightly higher ejecta masses.
Nevertheless, we recall that the number for the higher-mass ZAMS that we consider is much reduced due to the IMF.

{We then used the inverse transform sampling technique \citep{devroye86} that converts uniform random numbers between 0 and 1 into ZAMS masses that are distributed according to the desired IMF.}
See Appendix \ref{appendix_A} for details.
We generated ZAMS values within the progenitor mass range, which shows a greater number of lower-mass stars due to the steep negative exponent of the IMF, as expected.
Once the ZAMS mass was drawn, the ejecta mass, $M_{\rm ej}$, was computed via interpolation from the GENEVA model outputs and randomly assigned to each of the synthetic PWNe.
{As can be seen from Fig. \ref{fig: parameter_distribution}, the ejecta mass in this work ranges from 7 to 15 M$_{\odot}$ resulting from our adopted ZAMS mass range.}

Note that previous efforts (see, e.g., \citealt{ctagps24, fiori22}) have considered the ejecta mass to be distributed according to a Gaussian distribution centered at 13 $\rm M_{\odot}$ with $\sigma = 3\, \rm  M_{\odot}$, ranging from a minimum of 5$\, \rm  M_{\odot}$ to 20 $\, \rm  M_{\odot}$. {The latter high-end of the ejecta mass would in turn imply a very large ZAMS mass, with an unclear evolutionary path to a neutron star.}

In reality, the distribution of ejecta mass is barely constrained. Different models (or the same model with different parameters) result in a huge variation of the final masses.
For example, from \cite{sukhbold16} or \cite{renzo17}, it can be seen that the variation in the final mass considering different wind models and/or different efficiencies is extremely wide.
However, the ZAMS masses are better constrained, justifying our approach.

The impact of the distribution of the ejecta mass on the overall population, for example, in the characteristic luminosity–time ($L_0/L_{\rm ch}$–$\tau_0/t_{\rm ch}$) plane, which will be discussed later, is modest. 
For example, a uniform ejecta mass distribution ranging from 5$\, \rm M_{\odot}$ to 20$\, \rm M_{\odot}$ would not change the population distribution on the $L_0/L_{\rm ch}$–$\tau_0/t_{\rm ch}$ plane drastically from what is reported in this paper.
The spatial distribution of a single {random} realization of synthetic PWNe population of $1600$ sources on the backdrop of the Milky Way Galaxy is shown in Fig. \ref{fig: mwplot}, with each point color-coded with the corresponding ejecta mass, $M_{\rm ej}$.

\subsection{Pulsar characteristics}

Each simulated PWN is powered by a pulsar whose initial properties are drawn from observationally motivated distributions following \cite{bandiera23II}. 
The initial spin period was sampled from a Gaussian distribution with mean $\langle P_{0}\rangle = 100$ ms and standard deviation $\sigma_{P_{0}} = 80$ ms, truncated at a minimum value of 10 ms.
This choice lies intermediate between the distributions proposed by \cite{watters11}, \cite{johnston20}, and \cite{fgk06}, thereby sampling both the $\gamma$-ray and radio pulsar populations without overemphasizing either extreme, and remaining within the parameter space where associated PWNe are expected to form.
The pulsar initial magnetic field follows a lognormal distribution with mean $\langle \log_{10}(B_{0}/{\rm G}) \rangle = 12.3$ and dispersion $\sigma = 0.25$, similar to \cite{fgk06} and \cite{gullon15}, but with a cut at a minimum value of 10$^{12}$ G.

To calculate the initial period derivative, spin-down luminosity, and characteristic spin-down timescale, we employed the following relations (e.g., see \cite{gaensler06}).
The rotational evolution of a pulsar is primarily governed by the loss of its rotational kinetic energy due to electromagnetic torques. Assuming that the pulsar behaves as an isolated rotating magnetic dipole, the spin-down luminosity is given by
\begin{equation}\label{eq: edotrot}
    \dot{E}_{\rm rot} = -I \Omega \dot{\Omega},
\end{equation}
where \( \Omega = 2\pi/P \) is the angular velocity, \( \dot{\Omega} \) is its time derivative, and \( I \) is the moment of inertia of the neutron star. The evolution of \( \Omega \) follows a braking law of the form
\begin{equation}
    \dot{\Omega} = -K \Omega^n,
\end{equation}
where \( n \) is the braking index and \( K \) is a constant that encapsulates the dependence on the magnetic field strength, radius, inclination angle, and moment of inertia of the neutron star. For a pure magnetic dipole radiation in vacuum, the braking index is \( n = 3 \). In our simulations, we adopted a constant braking index of $n$ = 2.33, which corresponds to the secular braking index of Crab \citep{lyne15, horvath19}, to account for observed deviations from the ideal dipole model.
Rewriting the spin-down equation in terms of the period, \( P \), and integrating under the assumption that \( K \) and \( n \) remain constant yields the period evolution, we obtained
\begin{equation}
    P(t) = P_0 \left(1 + \frac{t}{\tau_0} \right)^{\frac{1}{n - 1}},
    \label{eq:P_t}
\end{equation}
where \( P_0 \) is the initial spin period.
Differentiating the above equation gives the time-dependent period derivative
\begin{equation}
    \dot{P}(t) = \frac{P_0}{\tau_0(n - 1)} \left(1 + \frac{t}{\tau_0} \right)^{\frac{2 - n}{n - 1}},
\end{equation} 
from which one can get the initial spin-down timescale, $\tau_0$:
\begin{equation}
    \tau_0 = \frac{P_0}{(n - 1) \dot{P}_0}\,.
\end{equation}
The relationship between the initial period, period derivative, and magnetic field can be obtained using the following expression:
\begin{equation}\label{eq: initial_B}
    B_0 \approx \sqrt{\frac{3 I c^n}{2(2\pi)^{(n-1)}R^{(n+3)}}} \times \sqrt{\dot{P_0}P_0^{(n-2)}} \ \mathrm{G}, \\
\end{equation}
which is a general formula with the dimension of the magnetic field, valid for any arbitrary braking index, $n$.
For example, for the magnetic dipole radiation in vacuum case ($n = 3$), with \( I = 10^{45} \ \mathrm{g \, cm^2} \) and $R = 10^6 \ \mathrm{cm}$, Eq. \ref{eq: initial_B} boils down to the well-known relation $B_0 \approx 3.2 \times 10^{19} \sqrt{P_0 \ \dot{P_0}} \ \rm G$. 

The age of a pulsar with present-day period \( P(t) \) and period derivative \( \dot{P}(t) \) can also be computed by integrating the spin-down law, leading to the expression
\begin{equation}
    t = \frac{P(t)}{(n - 1) \dot{P}(t)} \left[1 - \left(\frac{P_0}{P(t)}\right)^{n - 1} \right].
\end{equation}
The current age, period, and period derivative at the current age of the sources in the population are compatible with the formulation presented in this section. 

The initial spin-down luminosity of a pulsar can be calculated from its initial spin period, \( P_0 \), and period derivative, \( \dot{P}_0 \), using the relation
\begin{equation}
    L_0 =  4 \pi^2 I \frac{\dot{P}_0}{P_0^3} \ \mathrm{erg \, s^{-1}}.
\end{equation}
Once \( L_0 \) is determined, the time evolution of the spin-down luminosity can be described by

\begin{align}\label{eq: l(t)}
    L(t) &= 4 \pi^2 I \frac{\dot{P}(t)}{P(t)^3} \ \mathrm{erg \, s^{-1}} \nonumber \\
         &= L_0 \left(1 + \frac{t}{\tau_0} \right)^{-\frac{n+1}{n-1}} \ \mathrm{erg \, s^{-1}}\,,
\end{align}
where \( \tau_0 \) is the initial spin-down timescale and \( n \) is the braking index. 
It is worth noting that 
the true value of the braking index likely reflects a more complex and poorly constrained spin-down history. 
Given the limited number of measured braking indices available, we adopt this representative value throughout the paper; this also follows \citet{bandiera23II, bandiera23III}.

The histograms highlighting the distributions of the aforementioned parameters are shown in Fig. \ref{fig: parameter_distribution}. 
The sampling procedure employed in the parameter distribution histograms is discussed in Appendix \ref{appendix_B}.

\begin{figure*}[t!]
\centering
\raisebox{2.5mm}{\includegraphics[width=0.33\textwidth]{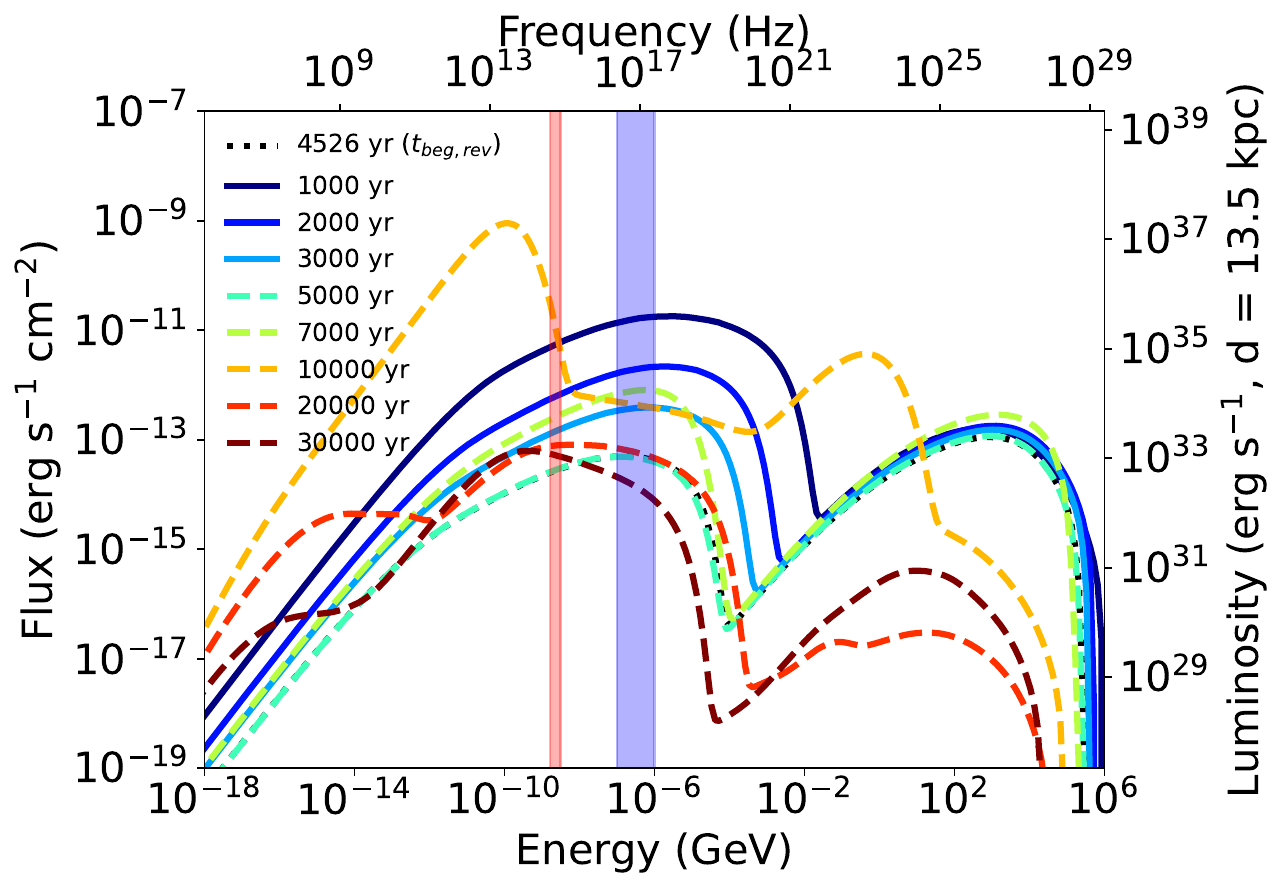}}
\includegraphics[width=0.33\textwidth]{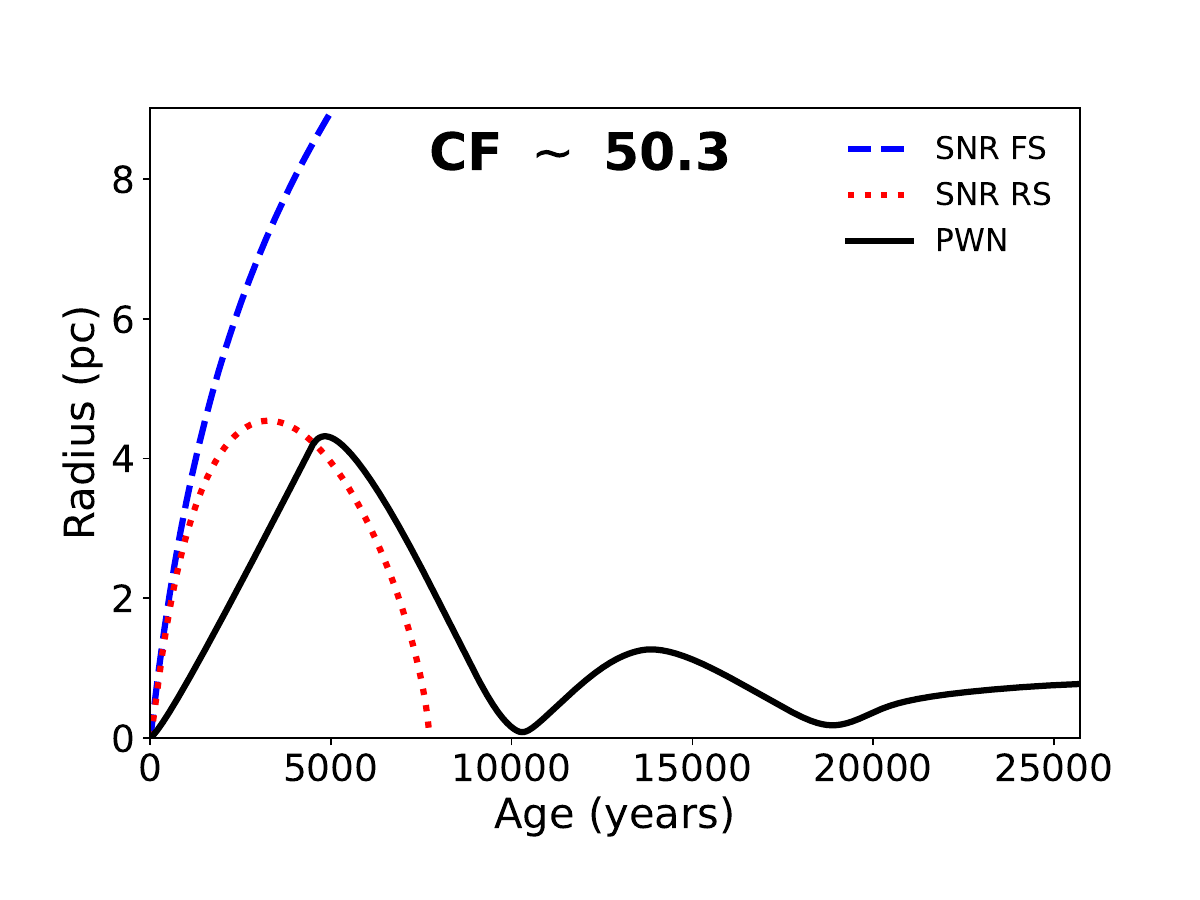}
\raisebox{2.5mm}{\includegraphics[width=0.28\textwidth]{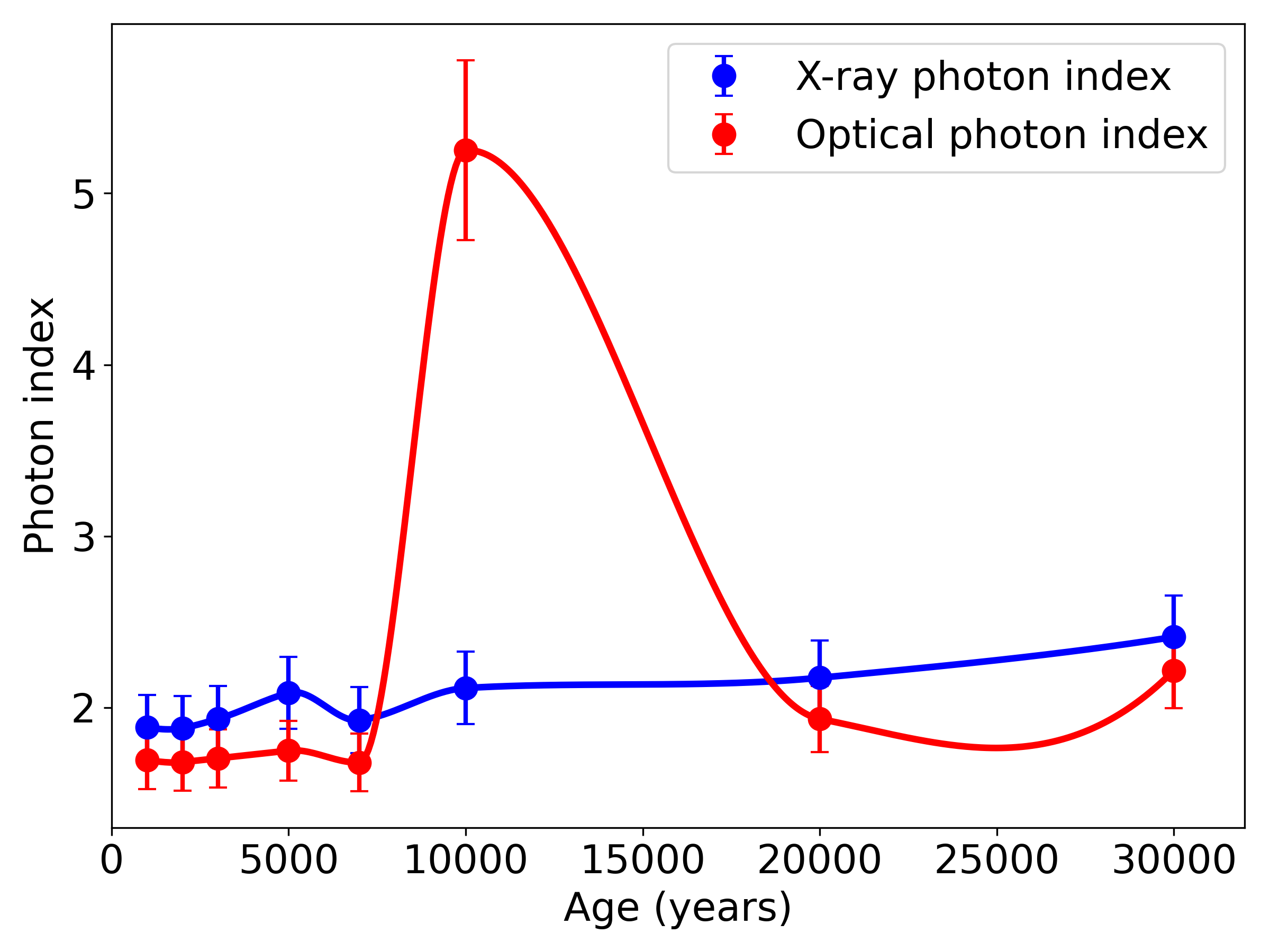}}
\raisebox{2.5mm}{\includegraphics[width=0.33\textwidth]{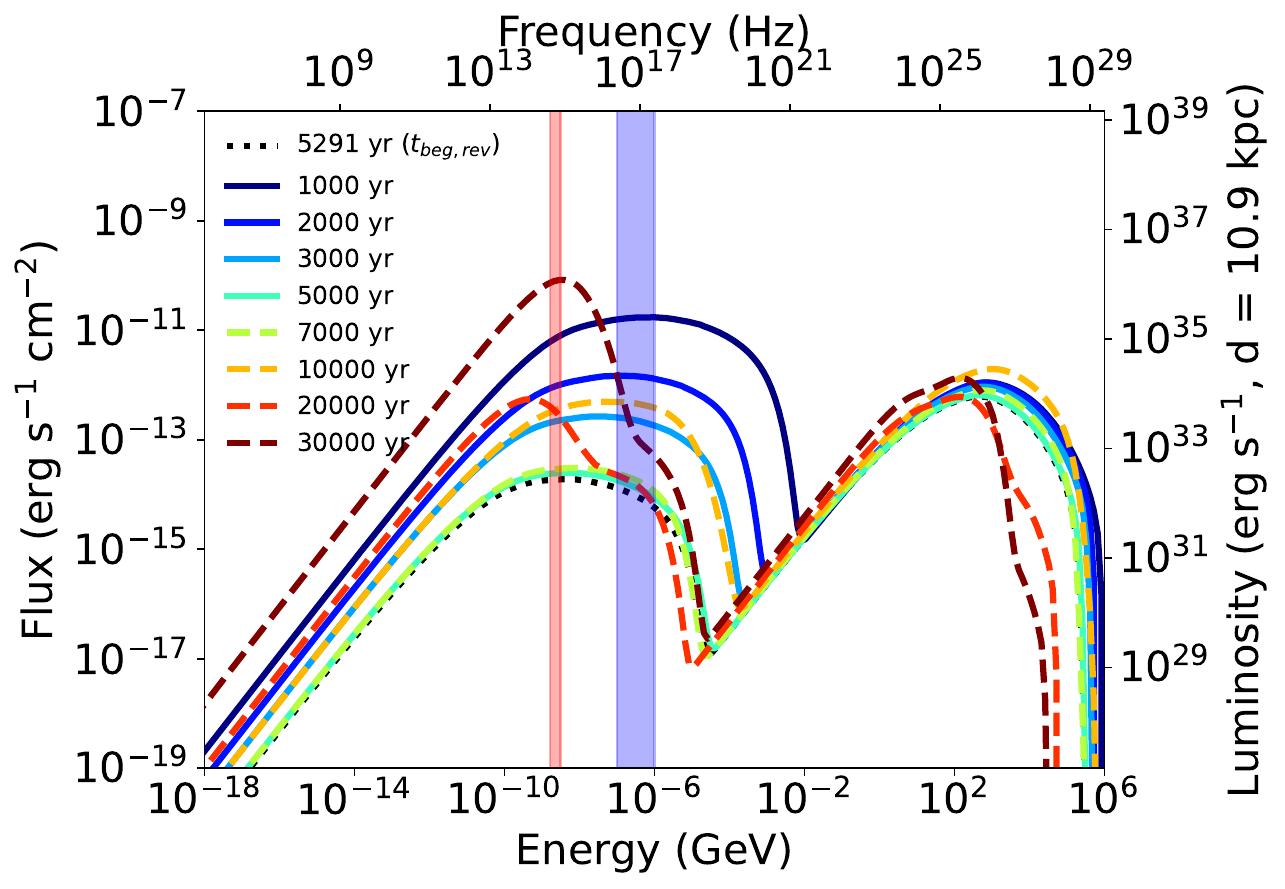}}
\includegraphics[width=0.33\textwidth]{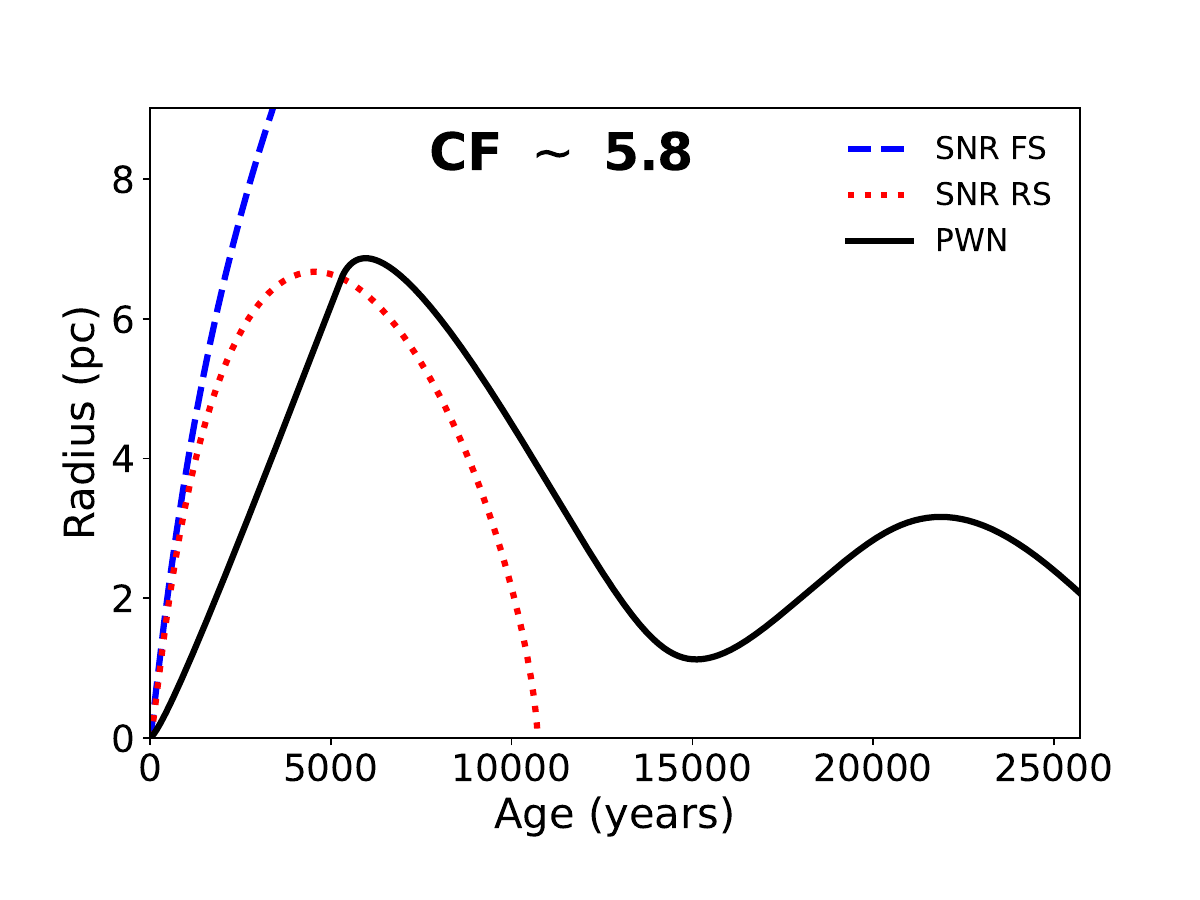}
\raisebox{2.5mm}{\includegraphics[width=0.28\textwidth]{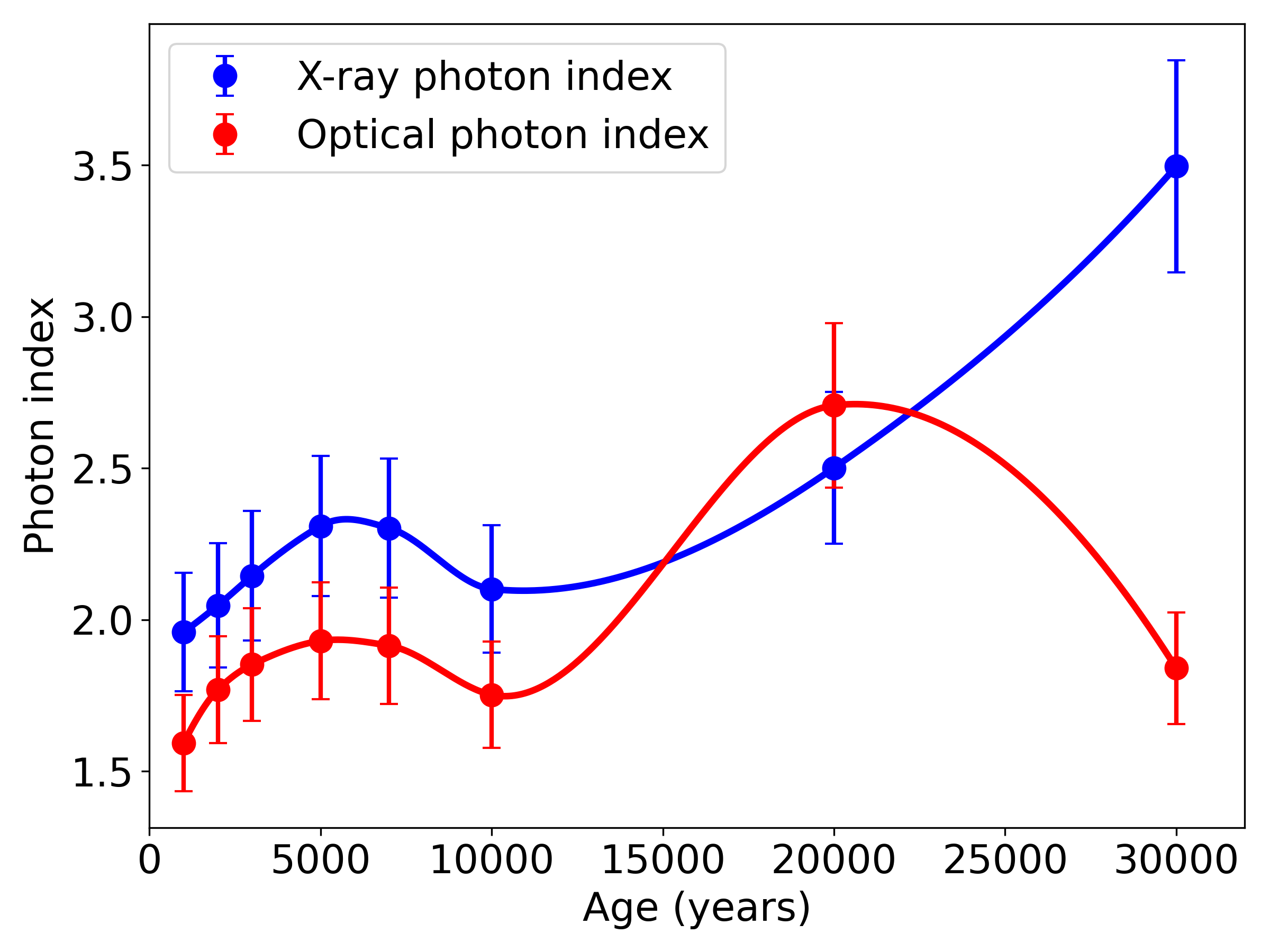}}
\caption{Results of the individual source simulations based on the CF are shown in this figure. The upper (lower) row corresponds to example 1 (example 2). The left panel in each row shows the evolution of the SED with time. The solid (dashed) lines correspond to the SED before (after) the reverberation, whereas the dotted line corresponds to the SED at the reverberation age ($t_{\rm beg, rev}$). The shaded red (blue) region signifies the energy range in which the optical (X-ray) photon index is calculated. The middle panel shows the time evolution of the radius of the PWN, SNR forward and reverse shocks. In the right panel, we show the time evolution of the optical (X-ray) photon index in red (blue) for both sources, while also providing 10$\%$ error bars for all photon index values.
}
\label{fig: individual}
\end{figure*}

\begin{table*}[htbp]
\centering
\caption{Parameters used for individual simulations of sources selected by their CF.}
\label{tab:individual_CF}
\begin{tabularx}{\textwidth}{XXXX}
\hline\hline
Definition & Parameter & Example 1 & Example 2 \\
\hline
Pulsar injection: & & & \\
\hline
Initial period & $P_0$ [s] & 0.10 & 0.07 \\
Initial period derivative & $\dot{P_0}$ [s s$^{-1}$] & $4.3 \times 10^{-12}$ & $1.5 \times 10^{-12}$ \\
Initial spin-down luminosity & $L_0$ [erg s$^{-1}$] & $1.6 \times 10^{38}$ & $1.6 \times 10^{38}$ \\
Initial spin-down age & $\tau_0$ [yr] & $5.7 \times 10^{2}$ & $1.1 \times 10^{3}$ \\
Final period & $P$ [s] & 1.81 & 0.95 \\
Final period derivative & $\dot{P}$ [s s$^{-1}$] & $1.6 \times 10^{-12}$ & $6.4 \times 10^{-13}$ \\
Current age & $t_{\rm final}$ [yr] & $2.5 \times 10^{4}$ & $3.4 \times 10^{4}$ \\
\hline
Environment: & & & \\
\hline
SN explosion energy & $E_{\rm SN}$ [$\times 10^{51}$ erg] & 0.98 & 0.72 \\
ISM density & $n_{\rm ISM}$ [cm$^{-3}$] & 1.09 & 0.22 \\
Ejected mass & $M_{\rm ej}$ [$M_{\odot}$] & 11.72 & 7.65 \\
Distance & $D$ [kpc] & 13.5 & 10.9 \\
\hline
Spectral model: & & & \\
\hline
Energy break & $\gamma_b$ & $3.4 \times 10^{5}$ & $4.7 \times 10^{6}$ \\
Low-energy index & $\alpha_1$ & 1.19 & 1.24 \\
High-energy index & $\alpha_2$ & 2.14 & 2.46 \\
Containment factor & $R_L/R_{\rm TS}$ & 0.45 & 0.86 \\
Magnetic fraction & $\eta$ & 0.0138 & 0.0014 \\
FIR energy density & $\omega_{\rm FIR}$ [eV cm$^{-3}$] & 1.92 & 0.99 \\
NIR energy density & $\omega_{\rm NIR}$ [eV cm$^{-3}$] & 1.38 & 2.76 \\
\hline
Derived quantities: & & & \\
\hline
Reverberation age & $t_{\rm beg,rev}$ [yr] & $4.5 \times 10^{3}$ & $5.2 \times 10^{3}$ \\
Compression factor & $R_{\rm max}/R_{\rm min}$ & 50.31 & 5.85 \\
\hline\hline
\end{tabularx}
\end{table*}

\subsection{Assumed model parameters}

Additional parameters relevant to the nebular emission were sampled from distributions motivated by previous studies \citep{fiori22,torres14}. 
These include the energy break in the injected particle spectrum {(sampled from a log-normal distribution, with $\log_{10}(\gamma_{\rm b})$ having mean 5.74 and standard deviation 0.37, which aptly covers the plausible $\gamma_b$ range from $5 \times 10^4$ to $1 \times 10^7$)}, low- and high-energy spectral indices (uniform distribution in $1.0 < \alpha_1 < 1.7$ and $2.0 < \alpha_2 < 2.7$, respectively), the containment factor (uniform distribution in $0.01 < R_{\rm L}/R_{\rm TS} < 0.9$, where $R_{\rm L}$ and $R_{\rm TS}$ are the Larmor radius and termination shock radius, respectively), and the magnetic fraction ({log-normal distribution with $\log_{10}(\eta)=-1.80$ and standard deviation $0.35$, with the additional condition $\eta<1$, to cover the plausible range from $10^{-3}$ to 0.2. This is also consistent with the typical magnetic fraction posited in \cite{torres14}}).
For each system, the local interstellar radiation field (ISRF) was estimated by interpolating tabulated near- and far-infrared (NIR and FIR) energy densities according to the Galactic position \citep{porter06}. 
To account for the higher-energy densities inferred from spectral modeling of observed PWNe \citep{torres14}, multiplicative correction factors, randomly selected from an uniform distribution ranging from 2 to 10, were applied to both NIR and FIR components, with fixed temperatures of $T_{\rm NIR} = 3000$ K and $T_{\rm FIR} = 70$ K.
{Such enhancements are physically plausible, since PWNe are commonly located in complex environments within the Galactic plane, near star-forming regions, heated dust, and molecular clouds, where the radiation field can be significantly stronger than the large-scale, azimuthally averaged ISRF implemented in \cite{porter06}.}
Altogether, this setup provides a statistically robust and physically motivated ensemble of PWNe, with initial conditions covering the plausible range of pulsar and its environmental properties. These synthetic systems form the basis for the subsequent dynamical and radiative evolution calculations, which are briefly discussed below.

\subsection{Evolving the population}

The evolution of each synthetic PWN was computed using the hybrid framework \texttt{TIDE+L}, which combines the time-dependent one-zone code \texttt{TIDE} \citep{martin12,torres14, martin16, martin22} with the Lagrangian reverberation modules introduced by \cite{bandiera20, bandiera21, bandiera23II, bandiera23III}. 
This approach addresses the limitations of thin-shell models, {with analytical formalism of the host SNR structure assumed a priori}, during the interaction between the PWN and SNR reverse shock, where nonlinear feedback dominates the dynamics and radiative outcome.
During the early free-expansion phase, the nebula evolves under the thin-shell approximation. 
However, during the reverberation stage, when the SNR reverse shock approaches the nebular radius, earlier models have overestimated the degree of compression.
{In particular, the treatment of the swept-up shell and the host SNR structure becomes especially important during the reverberation phase, since the reverse-shock-driven compression of the PWN is transmitted through the surrounding SNR structure. 
Unlike earlier thin-shell approaches, which prescribe the surrounding SNR structure a priori, \texttt{TIDE+L} evolves the HD structure of the SNR self-consistently together with the PWN, including the dynamical feedback of the shell itself. This provides a more realistic treatment of the pressure transfer during reverberation and avoids the overly abrupt or artificially strong compression that may arise in more simplified thin-shell prescriptions (see, e.g., Figs.~12 and 13 of \citealt{bandiera23II}).}
Later prescriptions yielded smoother evolution but could not follow the shell thickness or internal shocks. 
The Lagrangian scheme, as implemented in \cite{bandiera23II, bandiera23III}, overcomes these issues by tracking discrete ejecta shells, capturing interface thickening, delayed compression, and multiple shock reflections. 
This enables realistic modeling of the PWN-SNR dynamics and their observational signatures.

The radiative evolution is also coupled to the dynamics through the time-dependent particle distribution, accounting for synchrotron, IC, bremsstrahlung, synchrotron-self-Compton (SSC), and adiabatic losses. 
The injected spectrum, powered by the pulsar's spin-down luminosity, partitions the available energy between relativistic particles and the magnetic field according to a chosen magnetization parameter, from which the resulting non-thermal emission is subsequently computed.

\section{Results}
\label{results}

\subsection{PWN evolutionary stages}

In this work, we have divided the evolution of the PWN-SNR system into three phases: 
\begin{enumerate}
    \item The free expansion phase, in which the PWN is freely expanding within the unshocked SNR ejecta, 
    \item The reverberation compressing phase, in which the PWN is getting compressed by the reverse shock, but the minimum radius has not yet been reached, and 
    \item The reverberation post-compression phase, in which the first minimum has been reached, and the PWN is evolving further.

\end{enumerate}

Fig.~\ref{fig: piechart} presents a pie chart illustrating the mean fraction of the simulated PWNe currently in each evolutionary stage. 
The sampling procedure employed to estimate the mean count of sources in each evolutionary stage and to produce the pie chart is discussed in Appendix \ref{appendix_B}.
Note that for PWNe in post-compression, the evolution of PWNe radii is complex, and to have them all in a single category is a simplification, as subsequent compression or expansion events can, in principle, happen.
However, since these will likely be damped by instabilities or anisotropies in the medium, this global measure is enough for our aims.

In order to simplify the characterization of the PWN-SNR evolution and the reverberation phase properties, and following \cite{bandiera23II, bandiera23III}, we introduced the SNR characteristic radius ($R_{\rm ch}$), time ($t_{\rm ch}$), and luminosity ($L_{\rm ch}$). 
These characteristic scales are dependent on the values of $M_{\rm ej}$, $E_{\rm SN}$, and $\rho_{\rm ISM}$, and can be written
\begin{align}
R_{\rm ch} &= M_{\rm ej}^{1/3}\rho_{ISM}^{-1/3}, \\
t_{\rm ch} &= E_{\rm SN}^{-1/2} M_{\rm ej}^{5/6} \rho_{ISM}^{-1/3}, \\
L_{\rm ch} &= \frac{E_{\rm SN}}{t_{\rm ch}}
\end{align}
(see, e.g., \citet{truelove99}). 

{The combined evolution of the PWN--SNR system can be characterized by the ratios $\tau_0/t_{\rm ch}$ and $L_0/L_{\rm ch}$ throughout its evolution \citep{bandiera23II}, so that each source can be represented as a point in the $\tau_0/t_{\rm ch}$--$L_0/L_{\rm ch}$ plane. 
Although this parametrization is strictly valid only in the non-radiative
case and when the dimensionless ejecta structure is held fixed across all systems \citep{bandiera23II,bandiera23III},
it remains a useful approximate framework more generally.
In particular, it captures not only the pulsar birth properties, but also, through the adopted normalization, the influence of the SN explosion and the ambient medium, while reducing the number of explicit parameters needed to describe the evolution. 
Within this plane, the upper region is populated by energetic, long-lived pulsars, whose nebulae expand rapidly and undergo only mild reverberation, resembling Crab-like systems with comparatively large radii and weak compression. 
By contrast, the lower region corresponds to weak, rapidly fading pulsars, whose nebulae are more strongly affected by the SNR reverse shock, leading to stronger compression and more pronounced reverberation. 
Intermediate regions describe transitional cases, in which the PWN grows substantially before reverberation sets in, but still experiences a significant compression phase, resulting in moderate compression factors (CFs) and a delayed spectral evolution.}

\subsection{Population on the $L_0/L_{\rm ch}$--$\tau_0/t_{\rm ch}$ plane}

Fig. \ref{fig: isocontour} shows a single {random} realization of $1600$ synthetic PWNe on the log$_{10}(L_0/L_{\rm ch})$–log$_{10}(\tau_0/t_{\rm ch})$ plane. 
We further used the Gaussian kernel density estimation (KDE) to produce a smooth, continuous estimate of the density by using Gaussian (bell-shaped) kernels centered on each point.
This let us identify regions where sources are more densely clustered on the plane, and thus visualize the spatial density distribution of the population.
We also show ellipses encompassing the 10th, 50th, and 98th percentiles of the entire population, along with constant ratios of PWN-SNR energetics $L_0 \tau_0/E_{\rm SN}$.

We further investigated the dynamical compression experienced by PWNe by computing the CF, defined as the ratio between the PWN radius at the first local maximum and the subsequent local minimum in the evolution of the PWN radius, i.e., $R_{\mathrm{max}}/R_{\mathrm{min}}$ always computed at the first compression event.
For each simulated source, we identified these turning points in the radius-time profile and calculated the corresponding CF, capturing the degree of reverberation-induced contraction.
By mapping the CF as a color gradient for all points over the characteristic pulsar initial spin-down luminosity and timescale ( \( \log_{10}(L_0 / L_{\rm ch}) \) - \( \log_{10}(\tau_0 / \tau_{\rm ch}) \) ) plane, we visualize how reverberation strength varies across different pulsar configurations. 
This enabled the identification of regimes where strong compression is more likely, and offered a framework for comparing model predictions with observed PWNe properties.
The CF distribution on \( \log_{10}(L_0 / L_{\rm ch}) \) - \( \log_{10}(\tau_0 / \tau_{\rm ch}) \) plane is plotted in the right panel of Fig. \ref{fig: isocontour} along with 10th, 50th, and 98th percentiles enclosing ellipses and constant ratios of PWN-SNR energetics.  

\begin{figure}[t!]
\centering
\includegraphics[width=\columnwidth]{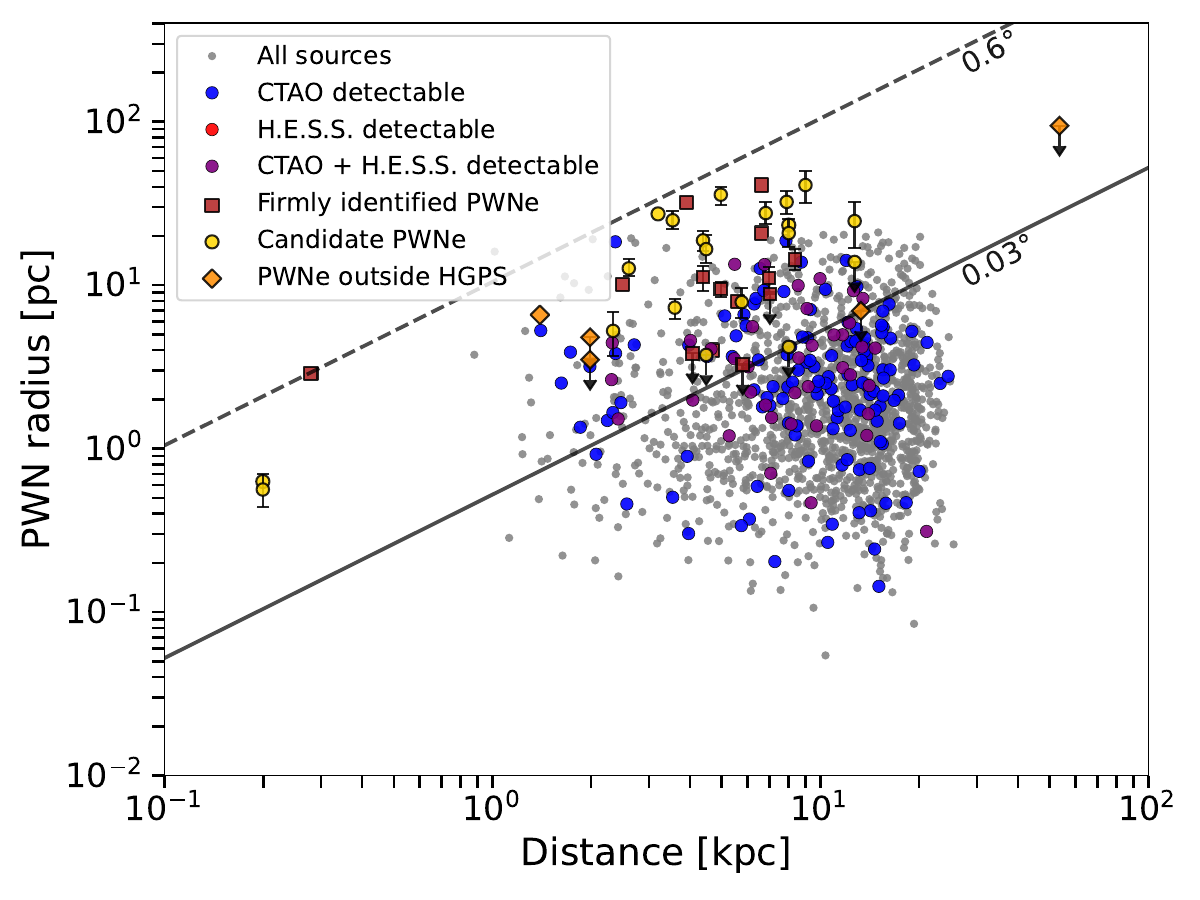}
\caption{Current PWN radius versus distance plot for a single {random} realization of the synthetic population containing $1600$ sources. {The red circles (if any) denote the sources that will be detectable by H.E.S.S., the blue circles represent the detectable sources by CTAO, and the purple circles denote sources that will be detectable by both}. The gray circles signify the rest of the population. The dashed and solid lines represent the maximum and minimum angular extension, respectively, as posited by \cite{hess18} from HGPS-detected PWNe. In the plot, firmly identified PWNe in the HGPS (firebrick squares), candidate HGPS PWNe (gold circles), and PWNe outside the HGPS (orange diamonds) are also shown. }
\label{fig: ppdot}
\end{figure}

\subsection{Individual examples of PWN evolution}

To show the time evolution of individual sources from our population, we selected two representative PWNe {within the 50$\%$ enclosing ellipse}, each characterized by a different CF, extracted from the distribution shown in Fig.~\ref{fig: isocontour} (their respective locations on the $L_0/L_{\mathrm{ch}}$–$\tau_0/t_{\mathrm{ch}}$ plane are indicated in the figure).
The corresponding input parameters for these sources are listed in Table~\ref{tab:individual_CF}.
In each case, we examined the multiwavelength (MWL) spectral energy distribution (SED) evolution over a series of snapshot ages to explore the effects of reverberation.
Additionally, we tracked the evolution of the SNR forward shock (FS), reverse shock (RS), and PWN radius to illustrate the onset and impact of the reverberation phase on the PWN structure.
As the system enters reverberation and undergoes compression, the magnetic field strengthens, leading to an increase in synchrotron luminosity and enhanced cooling. 
This results in a steepening of the SED, particularly in the optical to X-ray regime.
These individual PWN results are shown in Fig. \ref{fig: individual}.
To capture this behavior, we also present the time evolution of the photon index in the optical ($4 \times 10^{14} - 7.5 \times 10^{14}$ Hz) and soft X-ray band (0.1–1 keV) for both sources, computed at the same snapshot ages as above.

Following the onset of reverberation, the MWL SED of the sources exhibits complex, time-dependent evolution. 
In the high-compression case (Example 1; CF $\approx 50.3$, upper panel of Fig. \ref{fig: individual}), at the onset of reverberation, the PWN radius evolution clearly shows a sharp turnover and pronounced contraction of the PWN radius, marking a brief but powerful compression phase before gradual recovery. 
This leads to a sharp increase in synchrotron luminosity, peaking in the optical to X-ray regime, resulting from an amplified magnetic field. 
This is followed by a rapid steepening of the spectrum at higher energies as the amplified magnetic field drives strong radiative cooling. 
{The optical photon index initially hardens slightly at the onset of reverberation, as the particle population is rapidly reenergized during compression. 
This is followed by a brief but pronounced softening, caused by enhanced radiative losses in the amplified magnetic field during the early stages of reverberation. 
At later times, the optical spectrum hardens again, as continued particle injection and the cooling of higher-energy particles replenish the electron population radiating at optical energies.
The X-ray photon index shows a similar initial mild hardening at the beginning of compression, but subsequently softens more gradually as the evolution proceeds and the highest-energy particles continue to lose energy in the strengthened magnetic field.
In the high-CF case, the imprint of reverberation on the SED evolution is fast.}

In contrast, the low-compression case (Example 2; CF $\approx 5.8$, lower panel of Fig. \ref{fig: individual}) exhibits a milder and smoother evolution: the PWN compression due to reverberation is delayed, and further evolution is comparatively gradual.
Consequently, the magnetic field amplification, and thus the variation in the synchrotron component and cooling at the higher energies, is also comparatively delayed and gradual across time.
The overall SED evolution does not show a significant effect of reverberation long after the actual reverberation epoch ($t_{\rm beg, rev}$), owing to the slower response of the PWN to the crushing of the reverse shock.
{The optical photon index shows a mild hardening shortly after the onset of reverberation, followed by a delayed softening, and then hardens again as higher-energy particles cool and progressively repopulate the lower-energy, optical-emitting range.
By contrast, the X-ray photon index also exhibits an initial mild hardening, but then continues to soften gradually with time. 
This reflects the progressive depletion of the particles responsible for the X-ray emission, whose spectrum shifts to lower energies as the increasing magnetic field enhances synchrotron losses.
Note that in this case, the softening of the X-ray photon index sets in later, but is more pronounced and long-lived than in the high-CF case, reaching higher values and evolving over a more extended time interval.
}
Altogether, these examples illustrate that the magnitude of the CF primarily regulates the strength and duration of the luminosity amplification and spectral hardening or softening episodes during reverberation, highlighting how the diversity in CF across the population leads to markedly different observable signatures among middle-aged PWNe, which can be captured by X-ray instruments such as Chandra and XMM-Newton or detectors observing in the optical wavelengths.

\subsection{Source extension as a function of distance}\label{subsect: extension}

Fig. \ref{fig: ppdot} shows the distribution of the current age PWNe radii against the source distances for a single {random} realization of $1600$ synthetic PWNe, along with the maximum and minimum extension found in the H.E.S.S. Galactic Plane Survey (HGPS).
We also indicate the sources that would be detectable by CTAO and H.E.S.S. from our population in blue and red {(purple if detectable by both)}, respectively. 
{See Subsect. \ref{detectability} for details on the applied flux thresholds and observable sky coverage for each instrument.}

In Fig. \ref{fig: ppdot}, we also show the firmly detected and candidate PWNe by H.E.S.S., as well as PWNe outside HGPS.
{The extensions of the H.E.S.S.-detectable sources in our synthetic population remain broadly consistent with the range of extensions inferred from the HGPS data.}
Furthermore, our results indicate that CTAO is expected to detect a larger number of sources within the extension range (between 0.03 deg and 0.6 deg) posited by HGPS. 
Notably, most of the simulated sources do not exceed the maximum angular extension observed in the HGPS, while a significantly larger fraction are predicted to have radii smaller than the minimum range reported in the HGPS. 
{This is most probably the consequence of the treatment adopted here, which neglects mixing and enforces spherical symmetry in the coupled PWN--SNR evolution. 
On the one hand, this suppresses internal SNR anisotropies and mixing (see, e.g., \citealt{ferreira08}); on the other, it limits distortions at the PWN contact discontinuity, thereby naturally favoring more compact and regular nebular shapes.

Additionally, we adopt a braking index of $n=2.33$ rather than $n=3$, implying a faster decline of the pulsar spin-down luminosity than in the pure dipole case. 
Furthermore, note that in \cite{fiori22} the evolution of each PWN was artificially modified at the typical escape time from the SNR (see the insets of Fig.~3 in \citealt{fiori22}), which tends to produce larger late-time nebulae. 
More generally, our present setup does not include additional mechanisms that could increase the apparent extension of the tera-electronvolt-emitting region, such as a pulsar proper motion, which may generate displaced or elongated structures through the coexistence of relic and freshly injected particles, or the formation of extended pulsar halos beyond the nebula proper.

In addition, the simulations are followed only up to a finite maximum age, which may further limit the development of the largest systems. 
We therefore regard the present treatment as a reasonable compromise between physical realism and computational feasibility, while noting that it may still retain a residual bias toward underestimating the largest observed extensions.
Even without accounting for additional effects mentioned above that could further increase the sizes of PWNe, our simulations indicate that, given the angular resolution of CTAO, roughly half of the detectable sources should be spatially resolvable.

}

\begin{figure*}[t!]
\centering
\includegraphics[width=0.45\textwidth]{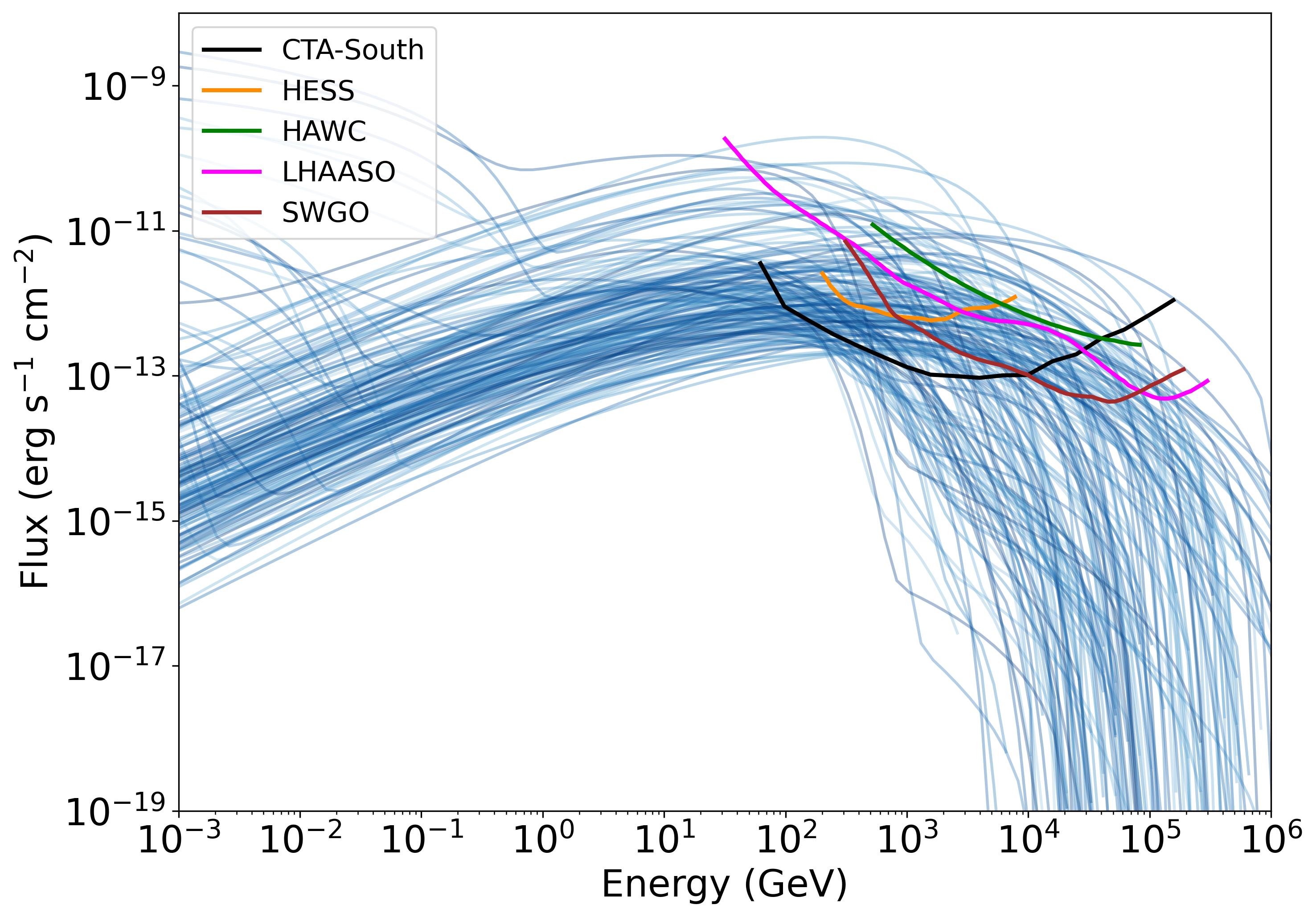}
\includegraphics[width=0.45\textwidth]{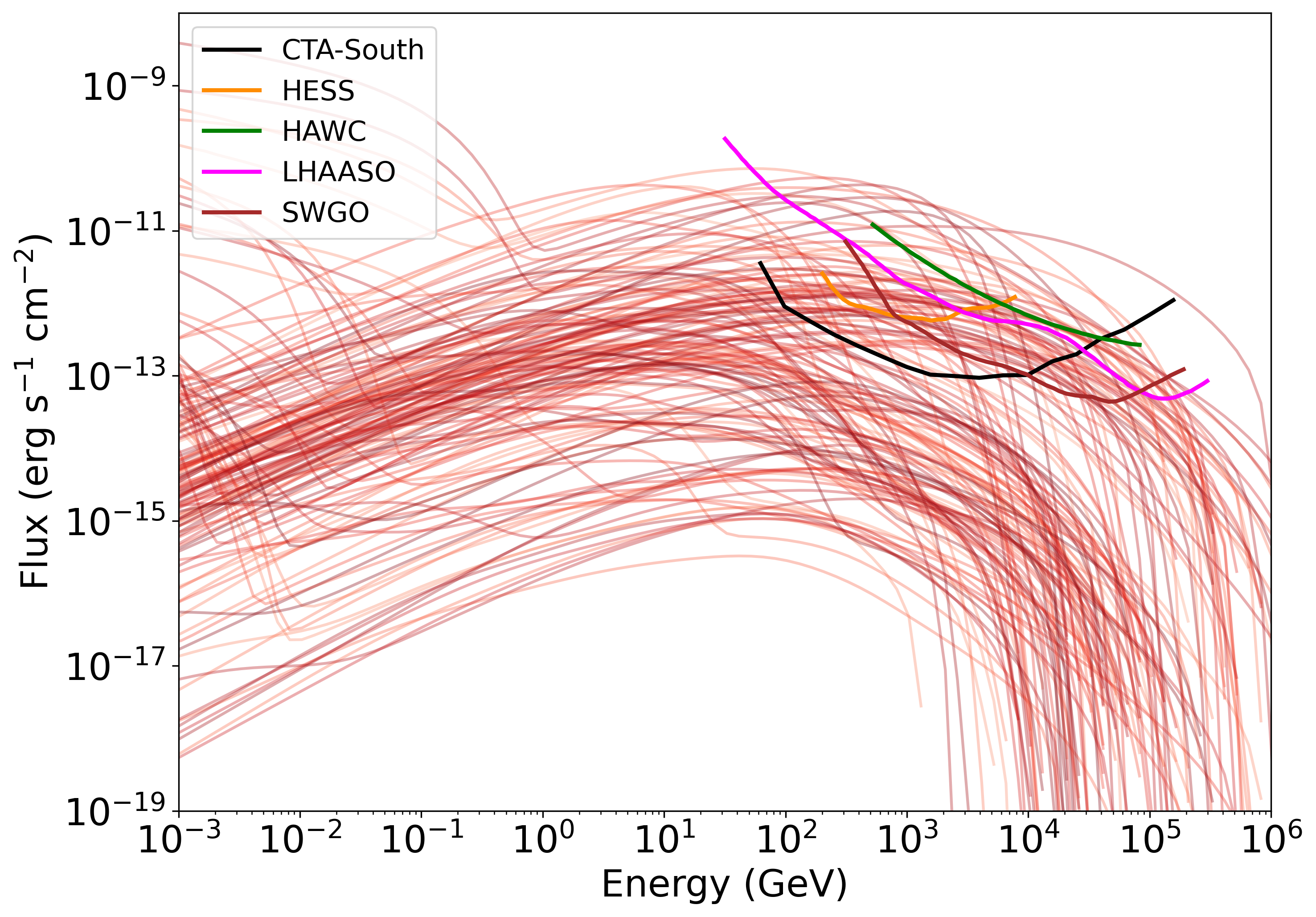}
\caption{Comparison of the SEDs of sources predicted to be detectable by CTAO for a single random realization of the simulated population taking the reverberation effects into consideration. For each source, the SED obtained with \texttt{TIDE+L} (left)  is compared with that of the corresponding source evolved with the simplified \texttt{TIDE} framework (right). The relevant instrumental sensitivity curves are also shown in both panels.}
\label{fig: TIDE_compare}
\end{figure*}

\begin{table}[t!]
\centering
\caption{Predicted detectability of simulated PWNe by current and future instruments for two modeling assumptions.}
\label{tab:detectability_all}
\begin{tabular}{lcc}
\hline\hline
\multicolumn{3}{c}{\bf \texttt{TIDE+L}} \\
\hline
Instrument & Detectable & Fraction (\%) \\
\hline
CTAO & 172 $\pm$ 7 & 10.7 \\
CTAO+ & 226 $\pm$ 8 & 14.1 \\
H.E.S.S. & 52 $\pm$ 4 & 3.2 \\
HAWC & 17 $\pm$ 2 & 1.0 \\
SWGO & 117 $\pm$ 6 & 7.3 \\
LHAASO WCDA & 35 $\pm$ 3 & 2.2 \\
LHAASO KM2A & 17 $\pm$ 2 & 1.0 \\
\hline\hline
\multicolumn{3}{c}{\bf \texttt{TIDE}} \\
\hline
Instrument & Detectable & Fraction (\%) \\
\hline
CTAO & 106 $\pm$ 6 & 6.6 \\
CTAO+ & 127 $\pm$ 6 & 7.9 \\
H.E.S.S. & 35 $\pm$ 3 & 2.2 \\
HAWC & 11 $\pm$ 2 & 0.7 \\
SWGO & 71 $\pm$ 5 & 4.5 \\
LHAASO WCDA & 23 $\pm$ 3 & 1.4 \\
LHAASO KM2A & 13 $\pm$ 2 & 0.8 \\
\hline
\end{tabular}

\tablefoot{%
The table reports the mean number of detectable sources across 1000 realizations, together with the corresponding $1\sigma$ uncertainty. In this table, ``fractions'' are computed as the mean number of sources that are detectable relative to a population of $1600$ PWNe.%
}
\end{table}

\subsection{Flux predictions and detectability}\label{detectability}

{In this section, we estimate the detectability of the simulated PWN population on a source-by-source basis, using the present-day gamma-ray output of each nebula together with the observational characteristics of each instrument. 
For every simulated source, the model provides a spectrum at its current evolutionary age, and this present-age spectrum was used to evaluate whether the source would be observable once the relevant sky coverage, flux sensitivity, and angular-extension effects were taken into account. 
The resampling procedure used to derive the predicted number of detections and the associated uncertainty is described in Appendix~\ref{appendix_B}.
We discuss below the detectability criteria for different instruments considered in this paper. 

CTAO: We assumed the alpha array configuration for CTAO.
We estimated detectability using the present-day integrated photon flux above 125~GeV. 
Only the simulated sources lying inside the adopted CTAO sky coverage region were considered further. 
In the implementation used here, the latitude coverage was taken to be $-6^\circ \le b \le 6^\circ$ \citep{ctagps19, ctagps24}, while the longitude dependence of the survey sensitivity was treated in five segments; namely, $[-60^\circ,60^\circ]$, $[60^\circ,150^\circ]$, $[150^\circ,-150^\circ]$, $[-150^\circ,-120^\circ]$, and $[-120^\circ,-60^\circ]$.
\footnote{Note that CTAO-South flux threshold within 1-10 TeV range is quoted in \cite{fiori22} to be 5 $\times 10^{-14} \ \rm erg \ cm^{-2} \ s^{-1}$. However, the flux threshold in the same range should be  $\sim 3 \times 10^{-14} \ \rm ph \ cm^{-2} \ s^{-1}$ \citep{ctagps24}, which should translate to $\sim 1 \times 10^{-13} \ \rm erg \ cm^{-2} \ s^{-1}$, about an order of magnitude larger.} 
To each of these regions, we assigned different point-source sensitivities above 125~GeV, i.e., 1.8, 2.7, 3.8, 3.1, and 2.6~mCrab, respectively, according to \cite{ctagps19}.  
\footnote{The corresponding photon-flux thresholds are obtained by scaling the integrated Crab flux, i.e., adopting \citep{aharonian06a},
\begin{equation}\label{eq: crab}
\frac{dN}{dE} = 3.76 \times 10^{-11}\, E_{\rm TeV}^{-2.39}\,
\exp\!\left(-\frac{E_{\rm TeV}}{14.3}\right)
\ \ {\rm ph\ cm^{-2}\ s^{-1}\ TeV^{-1}},
\end{equation}
and integrating this spectrum from 125~GeV up. 
Thus, 1~mCrab corresponds to $10^{-3}$ of this reference Crab flux (or 1 Crab Unit ($> 125 \ \rm GeV$)) for CTAO.
For the other instruments, the lower energy bound used to compute the corresponding integrated Crab flux is adjusted accordingly, while retaining the Crab spectrum given above.
We note that alternative spectral forms for the Crab Nebula have been adopted in the literature, such as a log-parabola in \cite{magic15} and a smoothly broken power law in \cite{hess24}. However, the resulting integrated Crab fluxes are broadly consistent and do not differ significantly. In this work, we adopt Eq.~\ref{eq: crab}, as it is the form used in \cite{hess18}, thereby enabling a direct one-to-one comparison.

}
{Note that the effective exposure times will depend on the details of CTAO survey implementation. An example can be seen in Tables 6.3 and 6.4 of \cite{ctagps19}.} 
Thus, the CTAO sensitivity depends explicitly on the location of the source within the survey footprint.

A second key ingredient is the treatment of source extension. 
For each PWN, we derived an angular size from its final modeled radius and distance, using the geometrical relation $\theta = \tan^{-1}(R_{\rm PWN}/d)$, and we used this directly as the characteristic source size ($\sigma_{\rm src}$) in the detectability calculation. 
Here, note that the adopted $\sigma_{src}$ is a lower limit to the real size of the PWNe, which neglects effects such as mixing and the pulsar proper motion, which will lead to more extended PWNe as discussed in Subsect. \ref{subsect: extension}.
The point-source threshold in the appropriate longitude segment was then degraded according to source extension following the prescription given in \cite{hess18},
\begin{equation}\label{eq: degrade}
F_{\min}(\sigma_{\rm src}) = F_{\min,0}\,\frac{\sqrt{\sigma_{\rm src}^2+\sigma_{\rm PSF}^2}}{\sigma_{\rm PSF}},
\end{equation}
where $\sigma_{\rm PSF}=0.07^\circ$ is the adopted CTA point-spread-function scale and $F_{\min,0}$ is the point-source sensitivity for the relevant longitude interval. 
In addition, we imposed an upper angular-size cut and excluded sources with $\sigma_{\rm src}>2^\circ$ \citep{martin22_pop}, since such very extended systems would fall outside the regime for which the adopted survey sensitivity can be meaningfully applied. 
Additionally, similar to \cite{martin22_pop}, we also considered an upgraded version of CTAO, dubbed CTAO+ in the paper, where the point-source sensitivity was improved by a factor of 2 compared to that presented in \cite{ctagps19}.  

H.E.S.S.: For H.E.S.S., we followed the same source-by-source approach, using the present-day integrated photon flux above 1~TeV for each simulated nebula. 
We restricted the source population to the adopted HGPS sky coverage; namely, $|b|<3^\circ$ and $l<65^\circ$ or $l>250^\circ$ \citep{hess18}. 
The point-source sensitivity was taken to be 10~mCrab above 1~TeV \citep{martin22_pop}, {corresponding to 50 h of observation time}. 
In this case, the mCrab conversion was obtained from the same adopted Crab spectrum (Eq. \ref{eq: crab}), but now integrated from 1~TeV upward. 

Similarly to CTAO, the source extension was explicitly taken into account. 
The angular size of each PWN was derived from its modeled final radius and distance, and the effective H.E.S.S. threshold was degraded according to Eq. \ref{eq: degrade}, where here we used $\sigma_{\rm PSF}=0.08^\circ$ \citep{martin22_pop}. 
We also imposed an upper angular-size cut of $\sigma_{\rm src} \leq 0.7^\circ$, such that more extended systems were excluded from the detectable sample. 

HAWC: For HAWC, we again applied the same source-by-source procedure using the present-day integrated photon flux above 1~TeV. 
To evaluate the sky visibility, the simulated Galactic coordinates were converted to equatorial declination, and only sources lying within the adopted HAWC field of view, $-20^\circ \leq \delta \leq 60^\circ$, were retained \citep{hawc20}. 
The point-source sensitivity was taken to be 50~mCrab above 1~TeV \citep{abey17}, {corresponding to 5 years of observation time}. 
As above, the mCrab conversion was obtained from the adopted Crab spectrum Eq. \ref{eq: crab}, here integrated from 1~TeV upward. 
Source extension was treated with the same prescription introduced above, with $\sigma_{\rm PSF}=0.3^\circ$ \citep{hawc13a} and an upper angular size cut at 2$^{\circ}$ (arbitrary) being adopted in this case. 

\begin{figure*}
    \centering
    \includegraphics[width=0.45\textwidth]{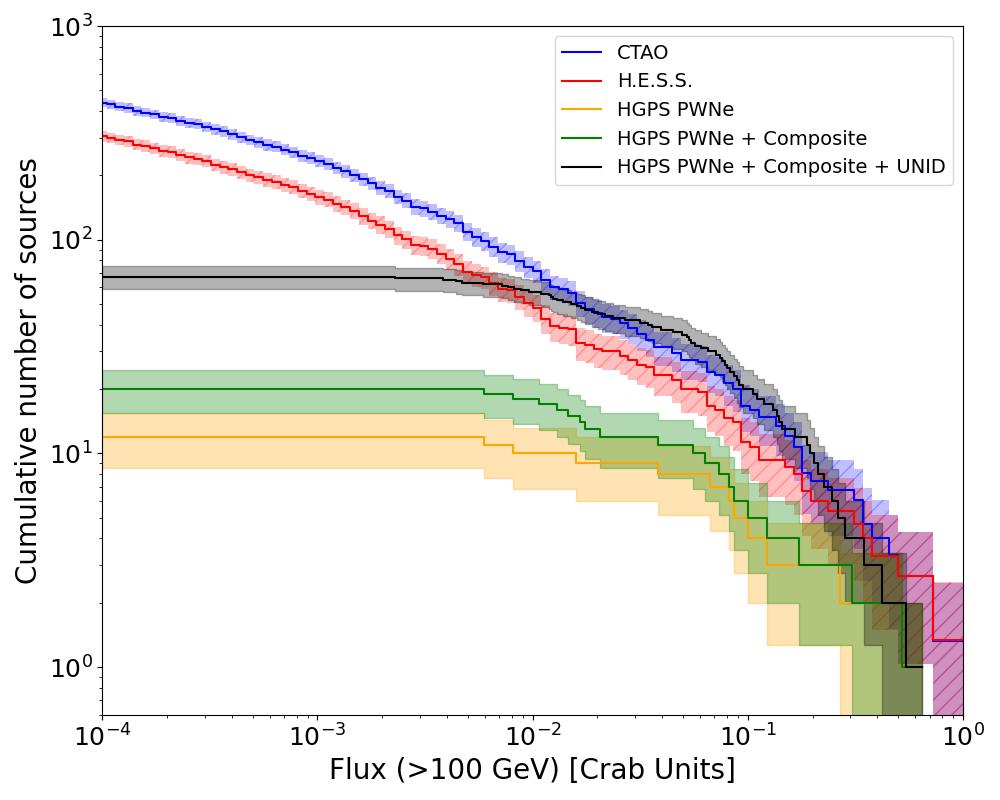}
    \includegraphics[width=0.45\textwidth]{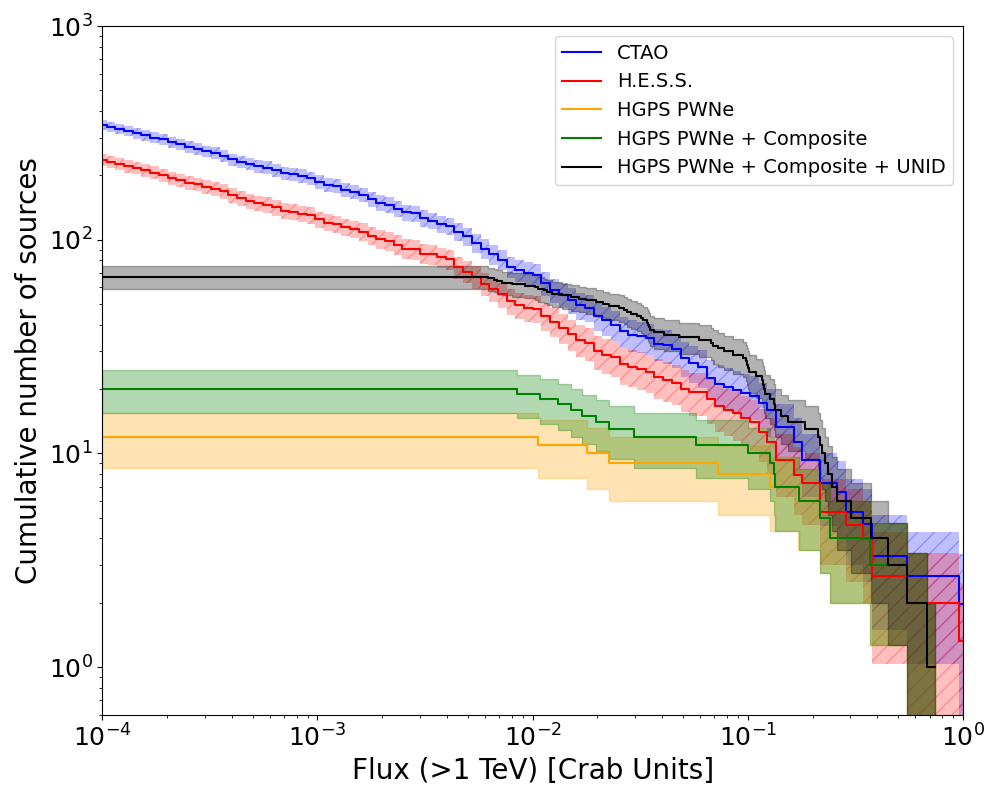}
    \caption{Mean cumulative flux distribution out of 1000 realizations of synthetic population containing $1600$ sources emitting above two specific energies, i.e., $>$ 100 GeV (left panel), and $>$ 1 TeV (right panel). The flux has been normalized with respect to 1 Crab unit. The lighter colored areas represent the errors of each curve. In both plots, the mean distribution of sources across the realizations of the synthetic population within the CTAO (in blue) and H.E.S.S. (in red) footprints are directly compared with the firmly identified PWNe (in orange), the same added with PWNe in a composite SNR (in green), and the sum of these two with the known HGPS unidentified sources (in black).}
    \label{fig: cumulative}
\end{figure*}

LHAASO: For LHAASO, we treated WCDA and KM2A within the same general framework, but unlike CTAO, H.E.S.S., and HAWC, the detectability was not evaluated through a single (or a series of) mCrab thresholds with an explicit PSF-based extension correction. 
Instead, for each source we computed the flux directly from the model spectrum at a fixed reference energy; namely, $F_3$ at 3~TeV for WCDA and $F_{50}$ at 50~TeV for KM2A. 
The source positions were converted from Galactic coordinates to equatorial declination, and only objects within the adopted LHAASO field of view, $-20^\circ \leq \delta \leq 80^\circ$ \citep{lhaaso24}, were considered. 
The source angular size was again inferred from the modeled final radius and distance and expressed as $r_{39}$.
{The observation time for LHAASO was considered to be the same as in \cite{lhaaso24}, i.e., 508 days for WCDA and 933 days for KM2A.}

For WCDA, detectability was determined by comparing $F_3$ with a size- and declination-dependent threshold surface reconstructed from the sensitivity curves given in \cite{lhaaso24}, using the $E^{-2.5}$ case. 
These curves provide the dependence of the threshold on declination for $r_{39}=0^\circ$ and $0.5^\circ$, and on $r_{39}$ for fixed declinations of $-15^\circ$ and $30^\circ$. 
We performed a two-dimensional interpolation in the $(\delta,r_{39})$ plane using the tabulated sensitivity grid.
A source was then counted as detectable by WCDA if its modeled $F_3$ exceeds the corresponding interpolated threshold. KM2A was treated analogously, but using the threshold surface for $F_{50}$ at 50~TeV, again based on the $E^{-2.5}$ sensitivity curves in declination and angular size. In this case, a source was counted as detectable when its modeled $F_{50}$ is larger than the corresponding interpolated KM2A threshold. 

SWGO: For SWGO, we followed an approach analogous to that adopted for LHAASO; namely, evaluating the source flux at a fixed reference energy rather than an integrated flux above threshold. 
In this case, we computed $F_1$ at 1~TeV directly from the resulting model spectra. 
The simulated Galactic coordinates were converted to equatorial declination, and only sources within the adopted SWGO sky coverage, $-70^\circ \leq \delta \leq 20^\circ$ \citep{scharrer25, swgo25}, were considered.
The point-source sensitivity was taken to be 5~mCrab at 1~TeV \citep{anguner24}, {corresponding to 5 years of observation time}, where the mCrab conversion, in this case, was defined from the differential Crab spectrum at 1~TeV (see Eq. \ref{eq: crab}). 
Source extension was then accounted for using the same prescription introduced above, adopting here $\sigma_{\rm PSF}=0.1^\circ$, and we additionally imposed an upper size cut of $\sigma_{\rm src}\leq 1^\circ$ \citep{scharrer25}. 

{Using the detectability criteria for different instruments discussed above and after applying the resampling procedure across 1000 realizations, we present the detectability statistics resulting from our synthetic population in Table \ref{tab:detectability_all}.}

\begin{table*}[t!]
\centering
\caption{Number of PWNe above instrument detection thresholds for each evolutionary phase.}
\label{tab:phase_detections}
\begin{tabular*}{\textwidth}{@{\extracolsep{\fill}}lcccccccc}
\hline\hline
Phase &
Count / Total &
Fraction (\%) &
CTAO &
H.E.S.S & 
HAWC &
SWGO &
WCDA &
KM2A \\
\hline

Free expansion & 153 $\pm$ 7 / 1600 & 9.5
& 49 $\pm$ 4 & 18 $\pm$ 2
& 5 $\pm$ 1 & 42 $\pm$ 4
& 12 $\pm$ 2 & 11 $\pm$ 2
\\

Compressing & 192 $\pm$ 8 / 1600 & 12
& 87 $\pm$ 5 & 32 $\pm$ 3
& 11 $\pm$ 2 & 66 $\pm$ 5
& 22 $\pm$ 3 & 6 $\pm$ 1
\\

Post-compression & 1255 $\pm$ 10 / 1600 & 78.4
& 36 $\pm$ 3 & 2 $\pm$ 1
& 1 $\pm$ 1 & 9 $\pm$ 2
& 1 $\pm$ 1 & 0 $\pm$ 0
\\

\hline
\end{tabular*}

\tablefoot{%
The table reports the mean number of detectable sources across 1000 realizations, together with the corresponding $1\sigma$ uncertainty. In this table, the ``fraction'' was computed as the mean number of sources in each evolutionary stage relative to a population of $1600$ PWNe.
}

\end{table*}

\subsubsection{Reverberation impact on detectability}

To further investigate how the detectability of PWNe depends on the treatment of reverberation, we compared the results obtained with the hybrid \texttt{TIDE+L} framework to those derived with \texttt{TIDE} \citep{martin12, martin16, martin22}, which adopts a simplified, one-zone, {purely} thin-shell description. 
For this comparison, we applied the same instrument-dependent detectability criteria and the same sampling procedure discussed in Appendix~\ref{appendix_B}, so that the differences can be directly attributed to the underlying dynamical treatment. 
The resulting predictions are summarized in Table~\ref{tab:detectability_all}. 
We find that the detectability is systematically reduced in the \texttt{TIDE} case across all instruments. 
{This behavior is consistent with the fact that the simplified treatment does not capture the reverberation phase with the same realism as \texttt{TIDE+L}, and may therefore overestimate the compression and the resulting magnetic field amplification, leading to excessively strong synchrotron losses, a reduced surviving high-energy particle population, resulting in a lower detectability fraction.}
The comparison thus reinforces the importance of modeling the reverberation phase appropriately when predicting the Galactic population of tera-electronvolt-emitting PWNe.

{To illustrate this effect more directly, we compare the broadband SEDs of individual sources between the two approaches. 
Specifically, we first identify the PWNe that are predicted to be detectable by CTAO in a single random realization of \texttt{TIDE+L} simulated population results, and then compare their broadband SEDs with those obtained for the corresponding sources in the \texttt{TIDE} setup. 
This allows the impact of the reverberation treatment to be assessed at the level of the emitted spectrum itself, rather than only through the total number of detectable objects. 
These comparisons are shown in Fig. \ref{fig: TIDE_compare} together with the relevant instrumental sensitivity curves. 
As can be seen there, the same sources that are detectable by CTAO in the \texttt{TIDE+L} simulation generally lie closer to, or even below, the sensitivity curves when modeled with \texttt{TIDE}. 
This demonstrates that a more realistic treatment of reverberation has a substantial effect on the predicted tera-electronvolt emission, and consequently on the inferred size of the Galactic tera-electronvolt PWN population.}

\subsubsection{Detectability as a function of PWN evolutionary stages}

We also performed the sampling method discussed in Appendix~\ref{appendix_B} for each evolutionary phase, and for each realization we applied the instrument-specific flux thresholds and sky-coverage cuts {as discussed above}. Table~\ref{tab:phase_detections} reports the mean detections averaged over all realizations, and the quoted uncertainties correspond to the sample $1\sigma$ standard deviation of the sampled ensemble. 
{A more realistic treatment of reverberation increases the expected number of detections because it avoids the overly strong compression that may arise in simplified treatments.
In such cases, the magnetic field can be overamplified during the compression phase, leading to excessively severe synchrotron losses that deplete the high-energy particle population and suppress the $\gamma$-ray emission.
By evolving the reverberation phase more self-consistently, \texttt{TIDE+L} yields a milder and more realistic compression, so that synchrotron cooling is not artificially overestimated and more sources remain detectable.
By contrast, sources in the post-compression stage are typically older, larger, and intrinsically dimmer, and therefore their detectability is generally lower.
In addition, the post-compression evolution is likely more sensitive to effects such as anisotropy, which may become important at late times but are not included in the present model.
For these reasons, only a small fraction of post-compression systems are predicted to be detectable.}

\subsection{Cumulative distribution of the population}

\begin{figure*}[t!]
\centering
\includegraphics[width=0.45\textwidth]{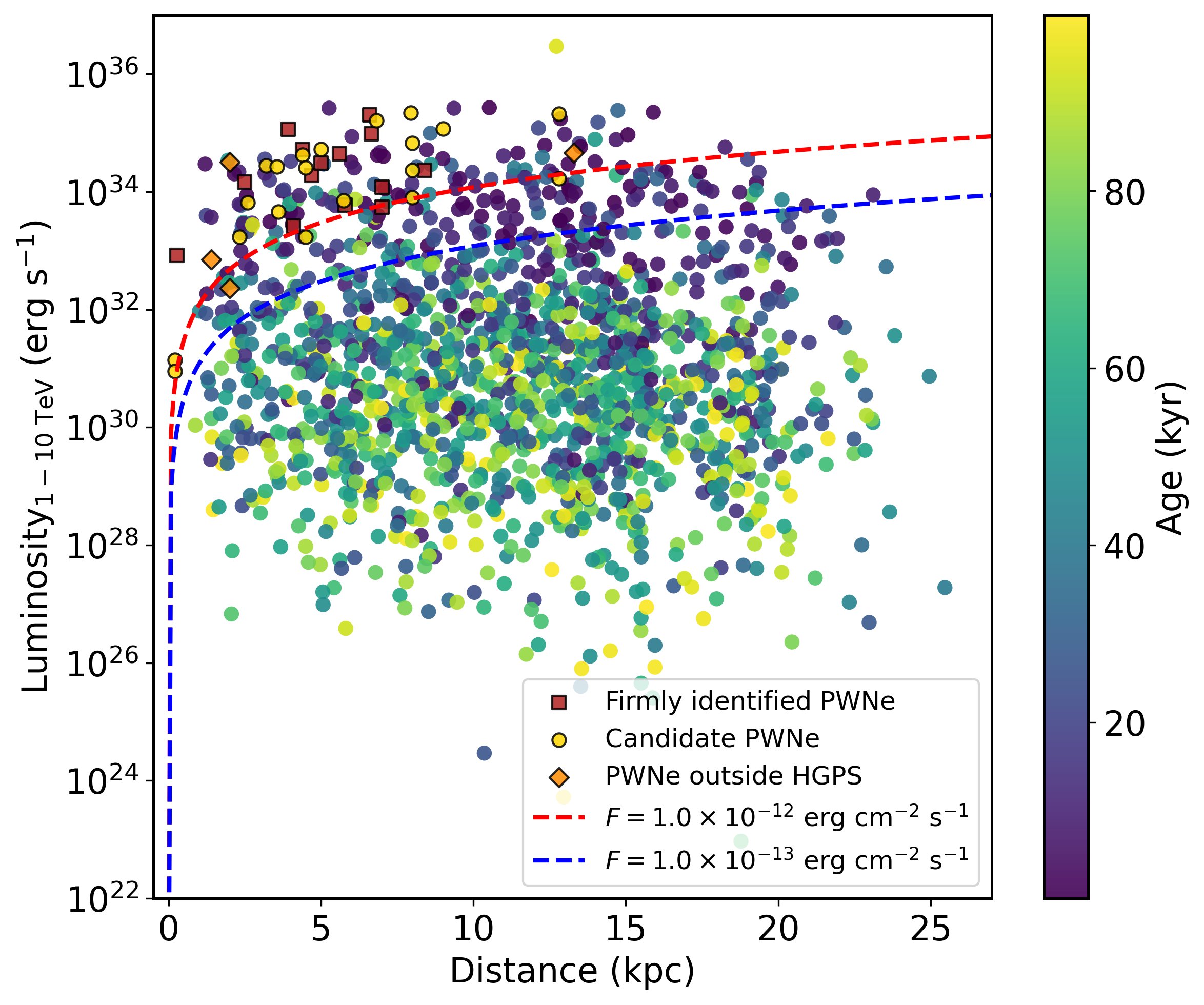}
\includegraphics[width=0.45\textwidth]{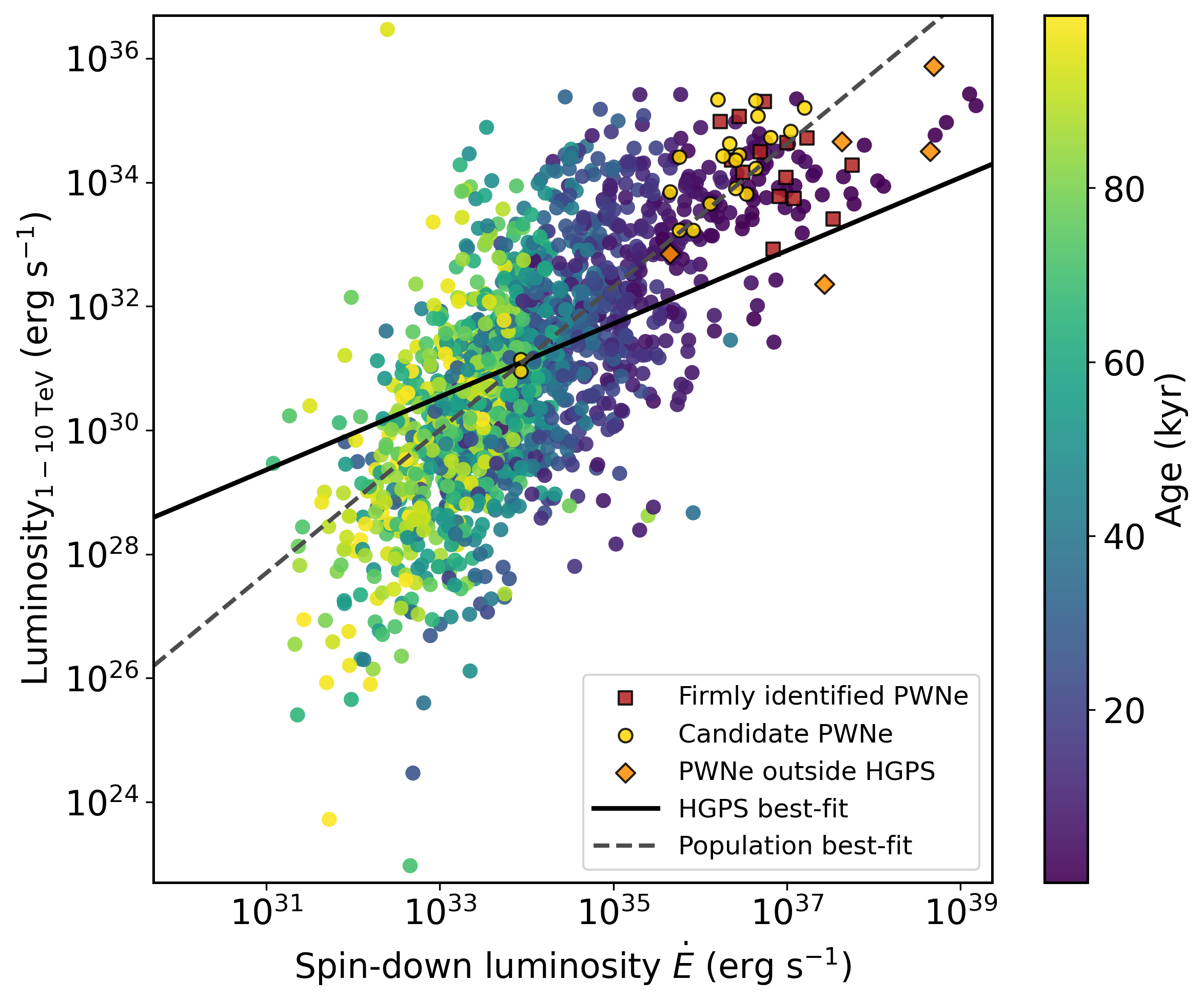}
\caption{Distribution of integrated $\gamma$-ray luminosities in the 1-10 TeV band with source distance (left panel) and spin-down power (right panel) for a single {random} realization of synthetic population containing $1600$ sources. In the left panel, the dashed red and blue curves signify the flux detection thresholds for H.E.S.S. \citep{hess18b} and CTAO \citep{ctagps24}, respectively. In the right panel, the solid line represents the slope corresponding to the $L_{1-10 \ \rm TeV} - \dot{E}$ relation, as obtained from \cite{hess18}, whereas the dashed gray line represents that calculated from our population. The population distributions in all panels have been color-coded based on the current age of the corresponding sources. In all of these plots, firmly identified PWNe in HGPS (firebrick squares), candidate HGPS PWNe (gold circles), and PWNe outside HGPS (orange diamonds) are also shown.}
\label{fig: aux}
\end{figure*}

To further assess the detectability of PWNe as a function of flux, we constructed cumulative distributions of sources, normalized to the Crab Nebula flux, above a certain energy threshold. 
For two chosen energy thresholds of $>$100~GeV and $>$1~TeV, we computed the expected integrated flux with respect to the Crab spectrum {as given in Eq. \ref{eq: crab}.}
Integrating this expression yielded the corresponding flux of 1~crab unit (CU) in photons\,cm$^{-2}$\,s$^{-1}$, with values of 1~CU ($>$100~GeV) $\approx 6.5 \times 10^{-10}\ {\rm photons\ cm^{-2}\ s^{-1}}$ and 1~CU ($>$1~TeV) $\approx 2.26 \times 10^{-11}\ {\rm photons\ cm^{-2}\ s^{-1}}$, which were then used to normalize the integrated PWNe fluxes above a corresponding energy threshold.

The resulting normalized fluxes were sorted and used to construct the cumulative distribution curves. 
Here as well, instead of relying on a single realization, we employed the sampling method discussed in Appendix \ref{appendix_B} to calculate the model cumulative distribution curves.
As discussed in Appendix \ref{appendix_B}, we used the mean cumulative distribution across all realizations of all the sources in the visible range of the respective instruments as the central model prediction in both cases, also including an error band computed as $\sqrt{N}$, where $N$ is the mean cumulative count for each flux threshold.
In both cases, we compared the model curves with HGPS results by overlaying cumulative distributions for different subsets: firmly identified PWNe, composite candidates, and unidentified (UNID) sources. 
This enabled a direct comparison between the simulated and observed source populations. 
For comparison, we selected all the sources from our population that are within H.E.S.S. and CTAO footprints, separately. 
The cumulative distribution plots are given in Fig.~\ref{fig: cumulative}.

By comparing both $>$ 100 GeV and $>$ 1 TeV distributions with the observables, it can be seen that both the simulated curves corresponding to CTAO and H.E.S.S. are consistent, within errors, with observational PWNe and PWNe+composite curves until $10^{-2}$ CU.
Below that value, the instrumental sensitivity decreases, resulting in the flattening of the observational curves.
The CTAO cumulative curve is consistent with the PWNe+composite+UNID curves at the highest flux  ($\gtrsim 0.2  \ CU$), indicating {that these unidentified sources at the highest fluxes have a strong possibility of being PWNe.}
The simulated curves also deviate slightly from the PWNe+composite+UNID curve between around {0.02 CU to 0.1 CU} in both cases (comparatively more pronounced in the $>$ 1 TeV case), which may indicate that {the SNRs, pulsar halos, or $\gamma$-ray binaries could contribute in that energy range.} 
The relative contributions of the middle-aged PWNe to the observable tera-electronvolt $\gamma$-ray sky will be directly testable with future CTAO observations.

\subsection{Tera-electronvolt luminosity distribution as a function of distance}

Fig. \ref{fig: aux} shows the integrated 1-10 TeV $\gamma$-ray luminosity for one {random} realization of $1600$ sources against distance (left panel) and spin-down power (right panel).
In the luminosity-distance plot, the red curve represents the H.E.S.S. detection threshold, whereas the blue curve corresponds to that of the CTAO. 
As expected, a vast amount of sources are outside the H.E.S.S. detection threshold, whereas a much larger fraction will be detectable by CTAO in the future.
{We further infer that about \(80\%\) of the sources lying above the H.E.S.S. and CTAO threshold have ages \(<40\) kyr.
Furthermore, about \(95\%\) of the sources above the H.E.S.S. threshold and \(85\%\) of those above the CTAO threshold are located within 15 kpc.
This trend is consistent with the presently firmly identified PWNe, which can all be found within $\sim 15$ kpc and have an age range within $\lesssim 40$ kyr.
We also find that only six sources above the H.E.S.S. threshold and nine above the CTAO threshold fall in the age range \(80 < t_{\rm age} < 100\) kyr. This suggests that comparatively old PWNe are likely to lie below the CTAO detection threshold. In turn, this may indicate that, if older systems are detected by CTAO, they are more likely to correspond to halo-like objects than to classical PWNe. }

Note that in previous literature, such as \cite{wilhelmi13}, it was posited that $\sim$ 300-600 PWNe would be detectable by CTA, especially in the CTA-I and CTA-D configurations.
However, the authors considered three PWNe as representative of the entire TeV PWNe population, and did not account for the time evolution of the PWNe sizes and luminosities, which is very complex, especially when reverberation is considered. 
Hence, previous estimations naturally lead to an overestimation, and at best, can be considered as an upper limit. 
Our estimates, while still in the hundreds of PWNe, are more conservative as well as more solidly based on PWN physics.

\subsection{Tera-electronvolt luminosity as a function of spin-down luminosity}

%
A correlation between X-ray luminosity and spin-down power was suggested by \cite{kargaltsev13}.
In the 0.1-100 GeV $\it Fermi$ band, a possible relation $L_{0.1-100 \ \rm GeV}$ with $\dot{E}$ can be inferred from \cite{Acero2013, Acharyya2025}, but data are too scarce to come to a meaningful conclusion.
In the tera-electronvolt energy range, the dependence was found to be $L_{1-10 \ \rm TeV} \propto \dot{E}^{0.59\pm0.21}$ \citep{hess18}, also with significant dispersion.

Here, the right panel of Fig. \ref{fig: aux} shows an apparent correlation between our integrated tera-electronvolt luminosity and the spin-down power.
When integrated $\gamma$-ray luminosity within 1-10 TeV is fit with a power law in logarithmic space, our population yields $L_{1-10 \ \rm TeV} \propto \dot{E}^{1.15}$.
The HGPS slope can be retrieved if we consider only the sources with spin-down power $\dot{E} > 5.9 \times 10^{35} \ \rm erg \ s^{-1}$, or sources with an age of fewer than {5.5 kyr}. 
This indicates that instrumental sensitivity limitations can introduce biases in the inferred slope.  
The luminosity threshold required to recover this correlation reflects the absence of survey-specific biases (e.g., flux- or extension-limited detectability) in our synthetic sample, while the younger age range suggests that the model predicts a more rapid decline or spectral softening of the tera-electronvolt emission with time -- likely driven by the evolution following reverberation -- than what is inferred from the HGPS data. 
Note that we do not report any uncertainties on both the estimation of slopes from our population, or after applying selection cuts to our population, where the inferred slope matches that posited by HGPS.
Given the large dispersion in the sample data points in which the slope is being fit in both cases, the uncertainties in the slopes do not bear much statistical meaning.
Nevertheless, this exercise demonstrates that the HGPS results can be plausibly reproduced, provided that appropriate selection cuts are applied.
For comparison, as before, we also show the firmly detected and candidate PWNe by H.E.S.S., as well as PWNe outside HGPS, in all plots of Fig. \ref{fig: aux}.

\section{Concluding remarks}\label{conclusion}

Given that PWNe are expected to dominate the tera-electronvolt $\gamma$-ray sky revealed by future observatories, accurately incorporating the physics governing their detectability is crucial. 
Our results are based on a more realistic treatment of reverberation compared to similar former approaches in the literature, implemented through the hybrid \texttt{TIDE+L} framework \citep{bandiera23III}.  
We find that neglecting reverberation dynamics leads to an underestimation of the $\gamma$-ray output, and thus to a lower fraction of detectable sources. 
Magnetic amplification, particle reheating, and subsequent losses induced by reverberation emerge as key ingredients shaping both luminosity distributions and SED diversity. 
Even though we caveat that reverberation is not a 1D process, and despite the improvement in our treatment, 
we are not addressing inhomogeneous compressions and the influence of the CSM in the evolution.
For timescales of tens of thousands of years, these can only be addressed by 3D simulations such as the ones in \cite{meyer25} using MHD, or \cite{olmi16}, \cite{porth14}, and \cite{kolb17} provided future advances in computing power allow for larger and longer simulations in relativistic magnetohydrodynamics (RMHD), which is currently unrealistic.

This study is not intending to reproduce the $P-\dot{P}$ diagram, as the latter is mostly populated by radio sources, much older than the ones we are interested in here.
In addition, we are using rotational losses for the pulsars, with a fixed braking index. 
Only a few values of the latter are known, and its range is vast.
Moreover, it can fluctuate and secularly vary across the pulsar evolution.
Although its influence on the evolution of the PWN is minor, it does significantly affect the pulsar period at sufficiently late times.
At such ages, other effects, such as magneto-thermal cooling, may be relevant to determine the $P-\dot{P}$ diagram.
These effects, never included in prior PWN studies, are also neglected in the present study as they are not expect to influence the population results, given the relatively young pulsar ages.
We shall explore in more detail if there is any possible impact of magneto-thermal cooling on the evolution of PWNe in a future work.

Qualitatively, the predicted number of {H.E.S.S. detectable PWNe in our population is consistent with} HGPS statistics \citep{hess18} {if the majority of the HGPS UNID sources are actually associated with pulsar-powered objects}, while CTAO is expected to increase detections by an order of magnitude.
{The updated CTAO+ configuration is expected to detect more than 200 PWNe in the Galaxy, in agreement with previous estimates.
We also predict that future facilities such as SWGO will play an important role in expanding and confirming the Galactic tera-electronvolt PWN population.
Our predicted LHAASO detectability is consistent with \cite{lhaaso24}, which reports 35 sources associated with pulsars. Of these, 24 are linked to pulsars younger than $10^5$~yr.
Moreover, 22 out of the 35 sources are classified as ultrahigh-energy emitters, with emission reaching or exceeding 50~TeV, again in line with our predictions.
}
The cumulative flux distributions suggest that many unidentified HGPS sources {at the highest fluxes} may correspond to evolved PWNe, while {also leaving room for} pulsar halos -- targets that future CTAO and SWGO surveys will directly test.

The derived correlation between integrated 1--10~TeV luminosity and pulsar spin-down power ($L_{\mathrm{1-10\,TeV}} \propto \dot{E}^{1.15}$) indicates that the well-known $L_{\gamma}$--$\dot{E}$ relation arises intrinsically from the coupled pulsar--nebula evolution rather than from survey biases. 
The spin-down luminosity threshold ($> 5.9 \times 10^{35} \ \rm erg \ s^{-1}$) needed to recover the HGPS survey trend underscores the absence of flux- or extension-limited biases in our synthetic sample, while younger systems ($t_{\rm age} < 5.5$ kyr) reproduce the steeper relation consistent with observations. 

We predict that most $\gamma$-ray-bright PWNe that are to be detected are likely younger than $\sim$ 40 kyr, while any older systems ($80 < t_{\rm age} < 100 \ kyr$) detected {are likely to} be associated with halo-like morphologies. 
This work hints at a natural evolutionary link between middle-aged PWNe and the emergence of extended tera-electronvolt halos around mature pulsars.
Individual case studies (Fig.~\ref{fig: individual}) further demonstrate how compression governs spectral hardening or softening and transient brightening, explaining the observed diversity of Galactic PWN SEDs.

Overall, our study provides a statistically robust picture of the Galactic PWNe population and its detectability, while incorporating a realistic formalism for reverberation. This approach bridges the connection between pulsar birth properties and survey observables, offering theoretical guidance for CTAO, LHAASO, and SWGO.
However, although comprehensive, our approach in this paper is not without caveats.
The escape of accelerated particles from the nebula into the host SNR, and subsequently into the surrounding medium, can modify the radiative signatures and extension of the system \citep{matin24, martin25}.
Moreover, a proper treatment of the CSM surrounding the PWN–SNR complex can significantly influence its dynamical evolution \citep{meyer24,meyer25}.
In addition, the total energy budget available from the pulsar spin-down includes a non-negligible pulsed component, which should be accounted for when evaluating the effective energy injection into the nebula.
We also did not take into account the possible escape of the pulsars from their host SNRs.
Another point that remains to be evaluated is the role of mixing, especially in the reverberation phase, and if there is a possibility of including it in one-zone models using an ad hoc recipe.

\begin{acknowledgements}
      We thank the anonymous reviewer for helpful suggestions and constructive criticism, which vastly improved the quality of this work. 
      This work has been supported by the grant PID2024-155316NB-I00 funded by MICIU /AEI /10.13039/501100011033 / FEDER, UE, and CSIC PIE 202350E189. This work is also partly supported by the Spanish program Unidad de Excelencia María de Maeztu CEX2020-001058-M, financed by MCIN/AEI/10.13039/501100011033, and by the MaX-CSIC Excellence Award MaX4-SOMMA-ICE, and also supported by MCIN with funding from European Union NextGeneration EU (PRTR-C17.I1).												
      ADS is supported by the grant Juan de la Cierva JDC2023-052168-I, funded by MCIU/AEI/10.13039/501100011033 and by the ESF+. 		
      D.F.T. acknowledges the T. D. Lee Institute, where part of this research was done, for hospitality.					
      BO and NB acknowledge partial support by the INAF Mini-grant HYPNOTIC87A: Hidden Young Pulsar Nebula Occupying The Inner Core of 87A and by the European Union – NextGenerationEU RRF M4C2 1.1 grant PRIN-MUR 2022TJW4EJ.

\end{acknowledgements}

%
   \bibliographystyle{aa} 
   \bibliography{aa.bib} 
%

\begin{appendix}

\section{Sampling stellar masses from a power-law initial mass function} \label{appendix_A}

\begin{figure*}
\centering
\raisebox{2.0mm}{\includegraphics[width=0.49\linewidth]{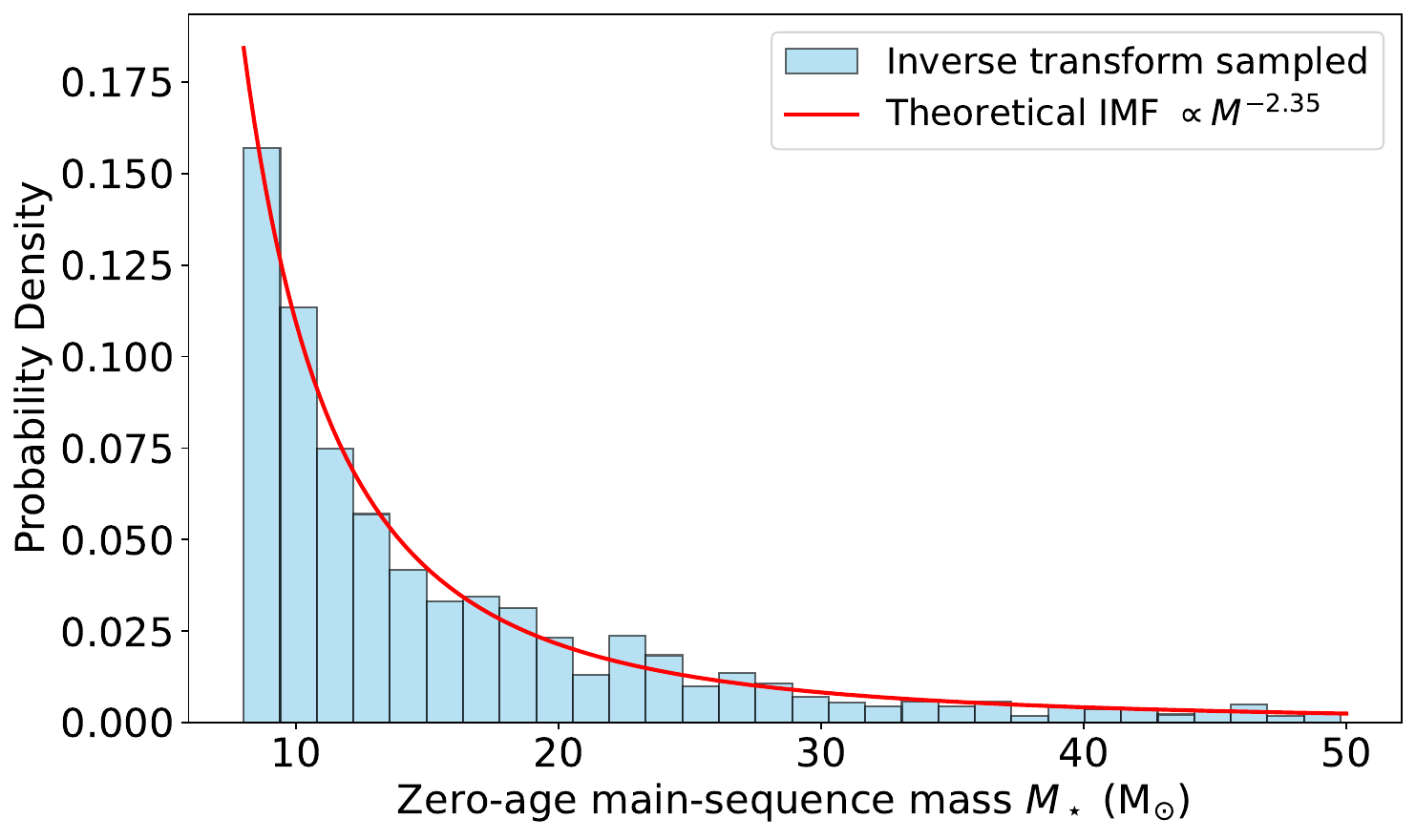}}
\includegraphics[width=0.47\linewidth]{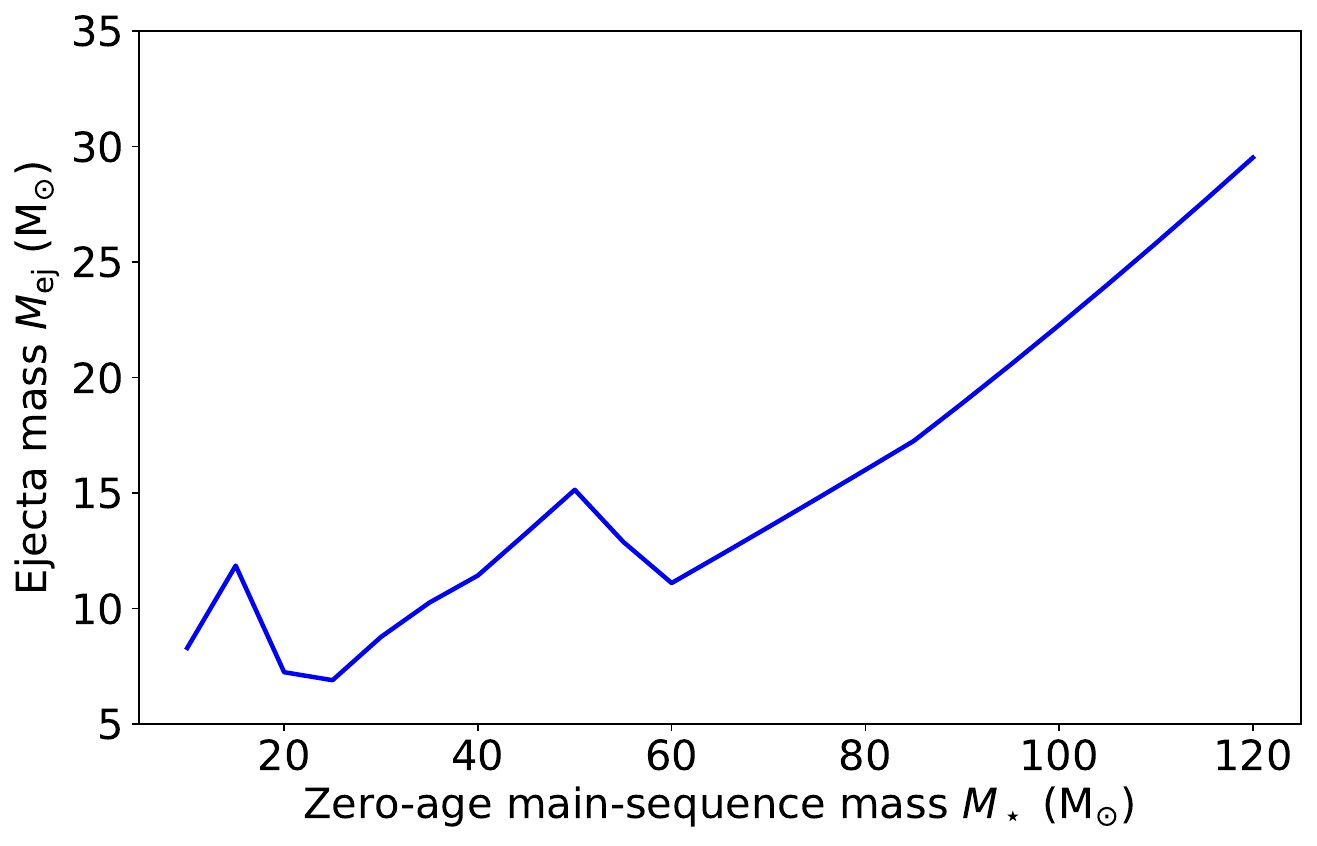}
\caption{The left panel shows the analytical IMF and 1600 randomly sampled ZAMS masses between 8-50 M$_{\odot}$ to ensure that we are sampling the IMF correctly. 
The right panel shows the relation between ZAMS mass and ejecta mass of nonrotating core-collapse SN progenitors at solar metallicity in the GENEVA library \citep{ekstrom12}. 
}
\label{fig: zams_mass}
\end{figure*}

To generate stellar masses following a prescribed IMF, we use inverse transform sampling \citep{devroye86}. This approach is well suited to power-law IMFs, since it allows direct sampling once the cumulative distribution function (CDF) is inverted.

We assume a power law probability density function of the form
\begin{equation}
p(M) = k\,M^{-\alpha},
\end{equation}
defined over the finite interval $[M_{\rm min}, M_{\rm max}]$, where the normalization constant $k$ satisfies
\begin{equation}
\int_{M_{\rm min}}^{M_{\rm max}} p(M)\,dM = 1.
\end{equation}
For $\alpha \neq 1$, this gives
\begin{equation}
k = \frac{1-\alpha}{M_{\rm max}^{\,1-\alpha} - M_{\rm min}^{\,1-\alpha}}.
\end{equation}

The corresponding cumulative distribution function is
\begin{equation}
F(M) = \frac{M^{1-\alpha} - M_{\rm min}^{\,1-\alpha}}
{M_{\rm max}^{\,1-\alpha} - M_{\rm min}^{\,1-\alpha}}.
\end{equation}
Setting $F(M)=U$, where $U \in [0,1]$ is a uniform random variable, and solving for $M$ yields
\begin{equation}
M(U)=
\left[
U\left(M_{\rm max}^{\,1-\alpha}-M_{\rm min}^{\,1-\alpha}\right)
+M_{\rm min}^{\,1-\alpha}
\right]^{1/(1-\alpha)}.
\end{equation}

This expression is then used to draw ZAMS masses distributed according to the desired power law IMF.
In Fig. \ref{fig: zams_mass}, we generated 1600 samples within the mass range [8, 50] $M_{\odot}$ , and the resulting distribution is compared with
the analytical IMF. As expected, the sample reproduces the power-law slope, with a greater number of lower-mass stars due to the steep negative exponent. 
Once the ZAMS mass is drawn, the ejecta mass $M_{\rm ej}$ is computed via interpolation from the GENEVA model outputs, which is also shown in Fig. \ref{fig: zams_mass}.

\section{Sampling procedure and statistical uncertainties}
\label{appendix_B}

To quantify the statistical uncertainty arising from the finite size of the synthetic Galactic PWN population, we adopt a common sampling framework that is applied consistently to all quantities discussed in this work. 
The procedure is designed to capture the intrinsic variance expected when drawing finite realizations from a larger synthetic parent population, and it is used to estimate uncertainties on parameter distributions, evolutionary phase fractions, detectability statistics, and cumulative flux distributions.

All sampling procedures described below are based on a parent population of $N_{\rm avail}$ simulated PWNe ($2400$ in this work). 
From this parent set, we repeatedly draw mock Galactic realizations by randomly selecting a fixed number $N_{\rm samp}=1600$ sources without replacement, ensuring that no source appears more than once in a given realization. 
Each realization therefore represents an equally plausible finite Galactic PWN population. 
This sub-sampling is repeated $N_{\rm real}=1000$ times, yielding an ensemble of independent realizations drawn from the same underlying synthetic population.

\subsection{Sampling of parameter distributions}\label{subappendix_B1}

For each parameter of interest, histogram bin edges are defined prior to the sampling, and the same bins are used for all realizations. 
In a given realization $k$, the sampled parameter values are binned to obtain a count $c_{k,j}$ in bin $j$. 
Repeating this process for all realizations produces an ensemble of counts $\{c_{k,j}\}$ that reflects fluctuations arising solely from finite-sample statistics.

The expected distribution is taken to be the mean count per bin,
\[
\bar{c}_j = \frac{1}{N_{\rm real}} \sum_{k=1}^{N_{\rm real}} c_{k,j},
\]
while the associated statistical uncertainty is estimated from the sample
standard deviation,
\[
\sigma_j = \sqrt{ \frac{1}{N_{\rm real}-1}
\sum_{k=1}^{N_{\rm real}} \left( c_{k,j} - \bar{c}_j \right)^2 }.
\]
In well-populated bins, the distribution of $c_{k,j}$ approaches a Gaussian by virtue of repeated random sampling, making $\sigma_j$ a meaningful $1\sigma$ estimate of the uncertainty. 
This procedure therefore yields a mean predicted distribution together with an intrinsic uncertainty arising from the finite size of the population.

\subsection{Sampling of evolutionary phase fractions}

Each simulated PWN is classified into one of three evolutionary phases: free expansion, reverberation--compressing, or reverberation--post--compression.
The sampling procedure described above is then applied to estimate the expected fraction of Galactic PWNe in each phase.

For each realization $k$, the sampled population is reduced to a set of counts $(c_{k,0}, c_{k,1}, c_{k,2})$, corresponding to the number of sources in each evolutionary phase. 
Across all realizations, this yields an ensemble $\{c_{k,i}\}$ for each phase $i$. The mean expected number of PWNe in phase $i$ is given by
\[
\bar{c}_i = \frac{1}{N_{\rm real}} \sum_{k=1}^{N_{\rm real}} c_{k,i},
\]
with an associated uncertainty
\[
\sigma_i = \sqrt{\frac{1}{N_{\rm real}-1}
\sum_{k=1}^{N_{\rm real}} \left( c_{k,i} - \bar{c}_i \right)^2 }.
\]
Because each count results from many independent random draws from a large population, the central limit theorem ensures that the distributions of $c_{k,i}$ are approximately Gaussian, allowing $\sigma_i$ to be interpreted as a $1\sigma$ uncertainty.

\subsection{Sampling for detectability}

For a given instrument considered, every simulated PWN is assigned an instrument-specific detectability flag based on its present-day model spectrum, sky position, and angular size. The sampling procedure is then applied to estimate the expected number of detectable sources and the associated sampling variance.

In each realization, we draw a population of size $N_{\rm samp}$ without replacement from the full eligible parent pool and evaluate the detectability criteria of the corresponding instrument for all sampled sources. Because the same realization is used to assess all instruments, correlated fluctuations between different observatories are naturally preserved. For a given instrument, the total number of detected sources in the $k$th realization is denoted by $n_{k,{\rm inst}}$. From the ensemble of $N_{\rm real}$ realizations, we compute the mean predicted number of detections as
\[
\bar{n}_{\rm inst} =
\frac{1}{N_{\rm real}}
\sum_{k=1}^{N_{\rm real}} n_{k,{\rm inst}},
\]
and the corresponding standard deviation as
\[
\sigma_{\rm inst} =
\sqrt{\frac{1}{N_{\rm real}-1}
\sum_{k=1}^{N_{\rm real}}
\left(n_{k,{\rm inst}}-\bar{n}_{\rm inst}\right)^2 }.
\]
These values are quoted as the expected detectability and its $1\sigma$ uncertainty. In practice, the distributions of $n_{k,{\rm inst}}$ are generally close to Gaussian, and their fit widths are consistent with the directly computed standard deviations, supporting the interpretation of $\sigma_{\rm inst}$ as the sampling uncertainty on the predicted number of detectable sources. In this paper, the resampling procedure is applied directly to the detectability flags, since the detectability analysis is carried out on a source-by-source basis using instrument-specific flux thresholds, sky-coverage constraints, and extension corrections.

\subsection{Detectability across evolutionary phases}

The detectability of PWNe as a function of evolutionary phase is estimated using the same sampling framework. 
In each realization, the sampled sources are first grouped according to their evolutionary stage. 
For each phase $i$ and instrument ``inst'', we compute the number of detectable systems $c^{(k)}_{i,{\rm inst}}$.

The mean number of detectable PWNe in phase $i$ for a given instrument is
\[
\bar{c}_{i,{\rm inst}} =
\frac{1}{N_{\rm real}} \sum_{k=1}^{N_{\rm real}} c^{(k)}_{i,{\rm inst}},
\]
with uncertainty
\[
\sigma_{i,{\rm inst}} =
\sqrt{\frac{1}{N_{\rm real}-1}
\sum_{k=1}^{N_{\rm real}}
\left(c^{(k)}_{i,{\rm inst}} -
\bar{c}_{i,{\rm inst}}\right)^2 }.
\]
As in the previous cases, the large number of realizations ensures that the resulting distributions are approximately Gaussian, and $\sigma_{i,{\rm inst}}$ provides a meaningful $1\sigma$ uncertainty.

\subsection{Sampling of cumulative flux distributions}

To compare model predictions with the observed HGPS cumulative source distributions, we construct ensembles of cumulative flux curves for CTAO and H.E.S.S. 
For each realization, instrument-specific sky coverage is applied to the sampled sources, and their integral fluxes above each flux threshold, for two different energy threshold cases ($E>1$~TeV or $E>100$~GeV) are computed and expressed in Crab Units.

For each realization $k$, sources are sorted by flux thresholds and used to construct a cumulative distribution $N_k(>F)$. 
Repeating this procedure for all realizations yields an ensemble of cumulative curves. 
The expected cumulative distribution is taken as the mean,
\[
\overline{N}(>F) = \frac{1}{N_{\rm real}} \sum_{k=1}^{N_{\rm real}} N_k(>F).
\]
The associated uncertainty is approximated as $\sqrt{\overline{N}(>F)}$, reflecting the typical Poisson-like variation and enabling a direct comparison with observational results.

\end{appendix}

\end{document}